\documentclass[smallextended]{svjour3}       
\smartqed  
\usepackage{graphicx}
\usepackage{epstopdf}
\usepackage{geometry}
\usepackage{graphicx,amssymb,amstext,amsmath}
\usepackage[justification=centering]{caption}

\usepackage{epsfig}
\usepackage{subfigure}
\usepackage{amsmath}
\usepackage{algorithmic}
\usepackage{graphics}
\usepackage{amssymb}
\usepackage{alltt}
\usepackage{float}
\usepackage{array}
\usepackage{xspace}
\usepackage{footmisc}
\usepackage{comment}
\usepackage{algorithm}
\usepackage{algorithmic}
\usepackage[misc]{ifsym}

\usepackage{multirow}
\usepackage{colortbl}

\begin{document}

\title{An Efficient Approach for Geo-Multimedia Cross-Modal Retrieval 
}
\subtitle{}


\author{Lei Zhu $^\dagger$$^{\natural}$        \and Jun Long $^\dagger$$^{\natural}$         \and
        Chengyuan Zhang $^\dagger$$^{\natural}$ \and
        Ruipeng Chen $^\dagger$$^{\natural}$     \and
        Xinpan Yuan $^\ddagger$     \and
        Zhan Yang $^\dagger$$^{\natural}$      \and
}


\institute{Lei Zhu \at
              \email{leizhu@csu.edu.cn}
           \and
           Jun Long \at
              \email{jlong@csu.edu.cn}           
           \and
           \Letter Chengyuan Zhang \at
              \email{cyzhang@csu.edu.cn}
           \and
           Ruipeng Chen \at
              \email{rpchen@csu.edu.cn}
           \and
           Xinpan Yuan \at
              \email{xpyuan@hut.edu.cn}
           \and
           Zhan Yang \at
              \email{zyang22@csu.edu.cn}
           \\
           $^\dagger$ School of Information Science and Engineering, Central South University, PR China \\
           $^{\natural}$ Big Data and Knowledge Engineering Institute, Central South University, PR China \\
           $^{\ddagger}$ School of Computer, Hunan University of Technology, China\\
}

\date{Received: date / Accepted: date}

\maketitle

\begin{abstract}
Due to the rapid development of mobile Internet techniques, cloud computation and popularity of online social networking and location-based services, massive amount of multimedia data with geographical information is generated and uploaded to the Internet. In this paper, we propose a novel type of cross-modal multimedia retrieval called geo-multimedia cross-modal retrieval which aims to search out a set of geo-multimedia objects based on geographical distance proximity and semantic similarity between different modalities. Previous studies for cross-modal retrieval and spatial keyword search cannot address this problem effectively because they do not consider multimedia data with geo-tags and do not focus on this type of query. In order to address this problem efficiently, we present the definition of $k$NN geo-multimedia cross-modal query at the first time and introduce relevant conceptions such as cross-modal semantic representation space. To bridge the semantic gap between different modalities, we propose a method named cross-modal semantic matching which contains two important component, i.e., CorrProj and LogsTran, which aims to construct a common semantic representation space for cross-modal semantic similarity measurement. Besides, we designed a framework based on deep learning techniques to implement common semantic representation space construction. In addition, a novel hybrid indexing structure named GMR-Tree combining geo-multimedia data and R-Tree is presented and a efficient $k$NN search algorithm called $k$GMCMS is designed. Comprehensive experimental evaluation on real and synthetic dataset clearly demonstrates that our solution outperforms the-state-of-the-art methods.
\end{abstract}

\section{Introduction}
\label{Intro}

Due to the rapid development of mobile Internet techniques, cloud computation and popularity of online social networking and search engine, massive amount of multimedia data is generated and uploaded to the Internet. For example, the most famous online social networking site, Facebook\footnote{https://facebook.com/}, has 1 billion 150 million users registered and 350 million photos uploaded daily as of November 2013. The total amount of images uploaded is 250 billion since its establishment. Twitter\footnote{http://www.twitter.com/} has more than 140 million users who posts 400 million tweets in the form of text and image all around the world. In China, the active users of Sina Weibo\footnote{https://weibo.com/} which is the largest micro-blog web site were 376 million as of September 2017, they post and share hundreds of thousands of texts, pictures or videos in this platform. More than 3.5 million new photos uploaded everyday in 2013 to Flickr\footnote{https://www.flickr.com/}, the most popular photo shared web site, and it had a total of 87 million registered users. For the video sharing service, YouTube\footnote{https://www.youtube.com/} shares more than 100 hours of videos every minutes as of the end of 2013. The number of independent users monthly in IQIYI\footnote{http://www.iqiyi.com/} which is the most popular video web service in China reached 230 million, and the total watch time monthly exceeded 42 billion minutes. The largest and most popular general reference work on the Internet, Wikipedia\footnote{https://www.wikipedia.org/}, comprises more than 40 million articles with pictures in 301 different languages. Unlike traditional structured data, these large-scale multimedia~\cite{DBLP:conf/mm/WangLWZ15} data has different modalities~\cite{DBLP:conf/mm/WangLWZZ14}, such as text, image, audio, video and even 3D objects in VR or MR fields. It is apparent that advanced online multimedia services with various modalities data not only bring great convenience for people in the daily life, but develop new requirements on multimedia data searching and sharing as well. On the other hand, the emergence of massive multi-modal data~\cite{NNLS2018} creates great challenges to data storage, mining and retrieval~\cite{DBLP:conf/ijcai/WangZWLFP16,DBLP:journals/corr/abs-1708-02288,DBLP:conf/pakdd/WangLZW14}. This necessitates the development of novel and efficient methods for multimedia data retrieval and processing.

As mentioned above, multimedia data contains several modalities~\cite{DBLP:journals/corr/BaltrusaitisAM17}, i.e., text, image, audio, video and 3D, which describes the world surrounding us, and each of them Corresponds to our each perception. For instance, our language can be written or spoken; natural scene can be represented by photos or videos; vocal signals can be recoded in audio files. In order to ulteriorly imitate human understanding of different modalities of data and enable search engine having the similar capacity, multi-modal and cross-modal representation and retrieval~\cite{Raymond2016Multimodal,DBLP:conf/aaai/CaoL0L17,DBLP:journals/tip/WangLWZZH15} problem had been proposed and it gets a lot of attention recently. Multi-modal and corss-modal retrieval~\cite{DBLP:conf/sigir/WangLWZZ15} involves feature extraction and fusion~\cite{DBLP:journals/mms/AtreyHEK10,Mcnamara2017Developing,DBLP:conf/cikm/WangLZ13,DBLP:journals/tnn/WangZWLZ17}, representation, semantic understanding, etc. And it is based on many techniques for monomodality retrieval.

Image is one of the most common modalities, and many image retrieval~\cite{DBLP:journals/mta/WuHZSW15} techniques support cross-modal retrieval. Content-based image retrieval (CBIR for short) is a hot issue in the multimedia area and lots of approaches have been proposed to improve precision and efficiency of image search. Several CBIR systems such as K-DIME~\cite{DBLP:journals/ieeemm/Bianchi-Berthouze03},IRMFRCAMF~\cite{DBLP:journals/remotesensing/LiZTZ16}and gMRBIR~\cite{DBLP:journals/prl/ChenWcYM16} have been proposed to develop advanced multimedia retrieval systems. Moreover, traditional feature extraction methods like scale-invariant feature transform (SIFT for short)~\cite{Lowe2002Object,Lowe2004Distinctive} and visual representation model such as bag-of-visual-words (BoVW for short)~\cite{Sivic2003A} are applied in cross-modal retrieval. Recently image recognition~\cite{DBLP:journals/pr/WuWGL18,TC2018} and retrieval based on CNN~\cite{DBLP:conf/iscas/LeCunKF10,Rawat2017Deep} is becoming a hot issue with the rise of deep learning techniques~\cite{DBLP:journals/tip/WangLWZ17}. For instance, ~\cite{Krizhevsky2012ImageNet} reports a quantum jump in image classification, which has the great improvement in performance in ImageNet large scale visual recognition challenge~\cite{Deng2009ImageNet}. Other works like~\cite{Hoffer2014Deep,Huang2017Learning,Melekhov2017Siamese} introduced serval new solutions for image search via deep learning. Another common modality is text, which exists over the Internet environment. Just like image retrieval, text search and understand plays an important role in both natural language processing and information retrieval communities. Many works using deep learning techniques, i.e., CNN~\cite{DBLP:conf/www/GuoYXYW17}, LSTM~\cite{DBLP:journals/corr/RocktaschelGHKB15,DBLP:conf/aaai/WanLGXPC16}, and siamese networks~\cite{DBLP:conf/emnlp/HeGL15} to develop novel solution for the problem of semantic textual similarity measurement~\cite{DBLP:conf/rep4nlp/YangYCKCPGSSK18,DBLP:journals/corr/WangMI16a} and retrieval~\cite{Liu2015Text}.

Dislike the monomodality retrieval above-mentioned, traditional cross-modal retrieval problem aims to find our objects with one modality by the query with another modality. For example, Searching an image on the Internet which can best demonstrate a given sentence or paragraph. Or finding an article or a poem in text which can describe a given photo. Example~\ref{ex:example1} is an example of traditional cross-modal retrieval which can depict cross-modal retrieval in a more specific way.

\begin{figure}[thb]
\newskip\subfigtoppskip \subfigtopskip = -0.1cm
\centering
\includegraphics[width=1.0\linewidth]{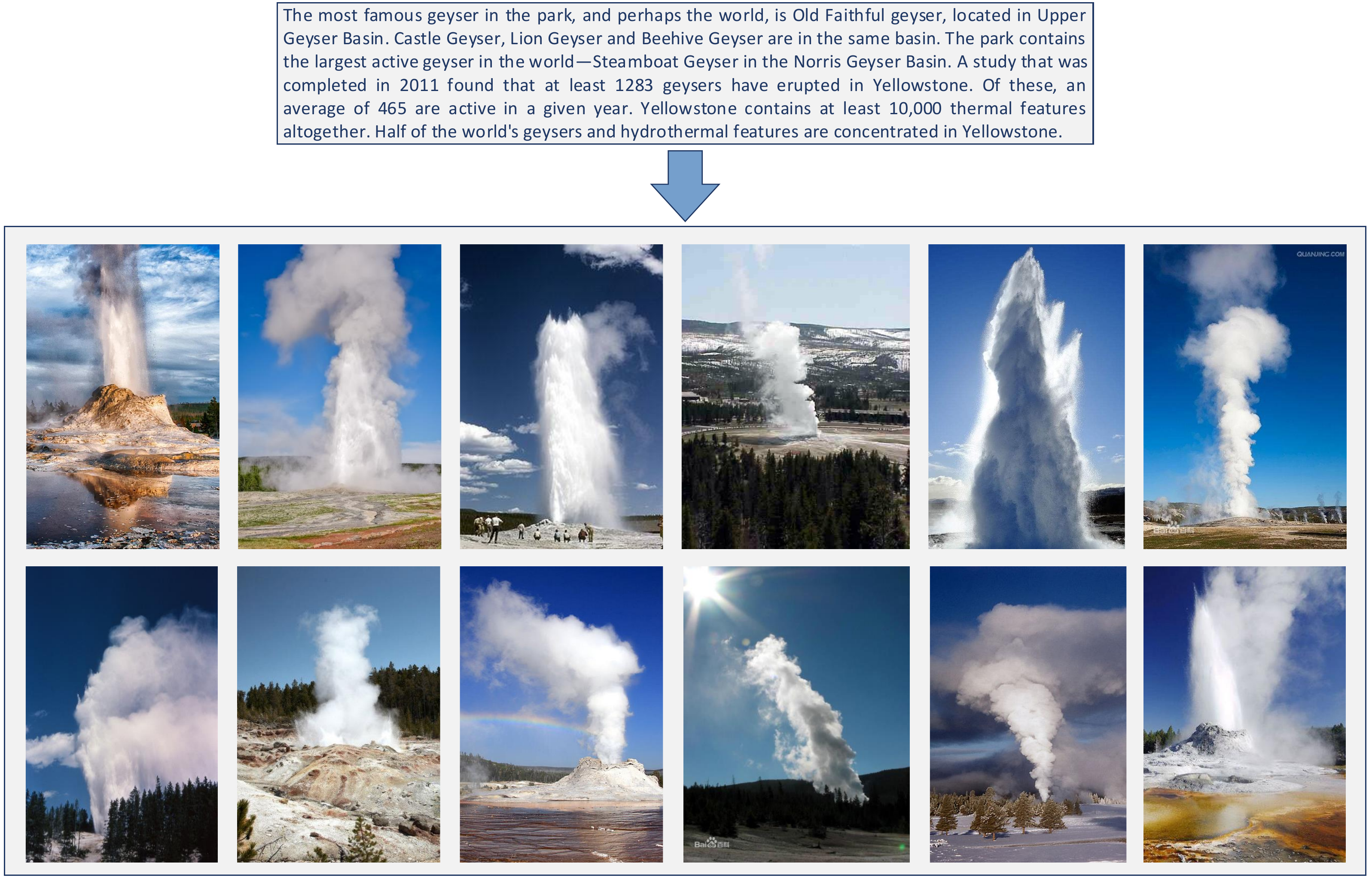}
\vspace{-1mm}
\caption{\small  An example of cross-modal retrieval }
\label{fig:fig1}
\end{figure}

\begin{example}
\label{ex:example1}
Figure.~\ref{fig:fig1} illustrates a typical example of cross-modal retrieval. An user need to find some pictures about famous geysers in Yellowstone National Park. She find a small paragraph of introduction of geysers in this park and put it into cross-modal retrieval system. This system then returns several images which are highly relevant to the input text by semantic analysis and similarity measurement. Unlike the keyword-based retrieval, cross-modal retrieval is based on understanding of multi-modal data and finding the semantic correlation.
\end{example}

As the locating techniques (e.g., GPS and gyroscope) and HD camera are applied widely in smart mobile devices such as smartphones and tablets, massive multimedia data with geo-tags, i.e., geo-images~\cite{DBLP:conf/dasfaa/ZhaoKSXWC15}, geo-texts and geo-videos can be conveniently collected and uploaded to the Internet. In Flickr, more and more photos are associated with geo-location information such as latitude and longitude. WeChat\footnote{https://weixin.qq.com/} is a very popular mobile application in which users can share geo-tagged multimedia data like texts, images and short videos with their friends. Other location-based services such as Google Places and Dianping combine texts, images and geo-location information to support spatial object query services, e.g., \emph{Where is the nearest seafood restaurant}, \emph{Which shop nearby sells this type of handbag}. Spatial textual or image queries is a hot spot in the spatial database community, which includes range queries~\cite{DBLP:conf/ssdbm/HariharanHLM07}, $k$NN queries~\cite{DBLP:conf/ssdbm/CaryWR10}, top-$k$ range queries~\cite{DBLP:journals/pvldb/CaoCJ10}, etc. It is concerned by lots of researches these days and several efficient indexing techniques like I$^3$~\cite{DBLP:conf/edbt/ZhangTT13}, KR$^*$-tree~\cite{DBLP:conf/ssdbm/HariharanHLM07}, IL-Quadtree~\cite{DBLP:conf/icde/ZhangZZL13,DBLP:journals/tkde/ZhangZZL16}, IR-tree~\cite{DBLP:journals/tkde/LiLZLLW11} and its variations~\cite{DBLP:journals/pvldb/CongJW09}, WIR-tree~\cite{DBLP:journals/tkde/WuYCJ12}, etc. have been proposed to improve performance of the system.

\noindent\textbf{Motivation}. It is a pity that although traditional spatial keyword or image queries have been particularly well studied, but they just consider monomodality during the retrieval processing. That means these approaches cannot be applicable to the cross-modal retrieval. On the other hand, previous studies of traditional multi-modal and cross-modal retrieval just concentrating on feature extraction and semantics correlation between different modalities, and cross-modal similarity measurement. However, they do not consider the geographical location information tagged with multimedia data. Undoubtedly, geographical location is another significant information for supporting advanced search engines and location-based services. To the best of our knowledge, there is no one who has paid attention on the problem of geo-multimedia cross-modal retrieval at present. In order to describe this new problem clearly, a motivating example is introduced below, in which both the cross-modal search and geographical distance proximity are considered.

\begin{figure}[thb]
\newskip\subfigtoppskip \subfigtopskip = -0.1cm
\centering
\includegraphics[width=1.0\linewidth]{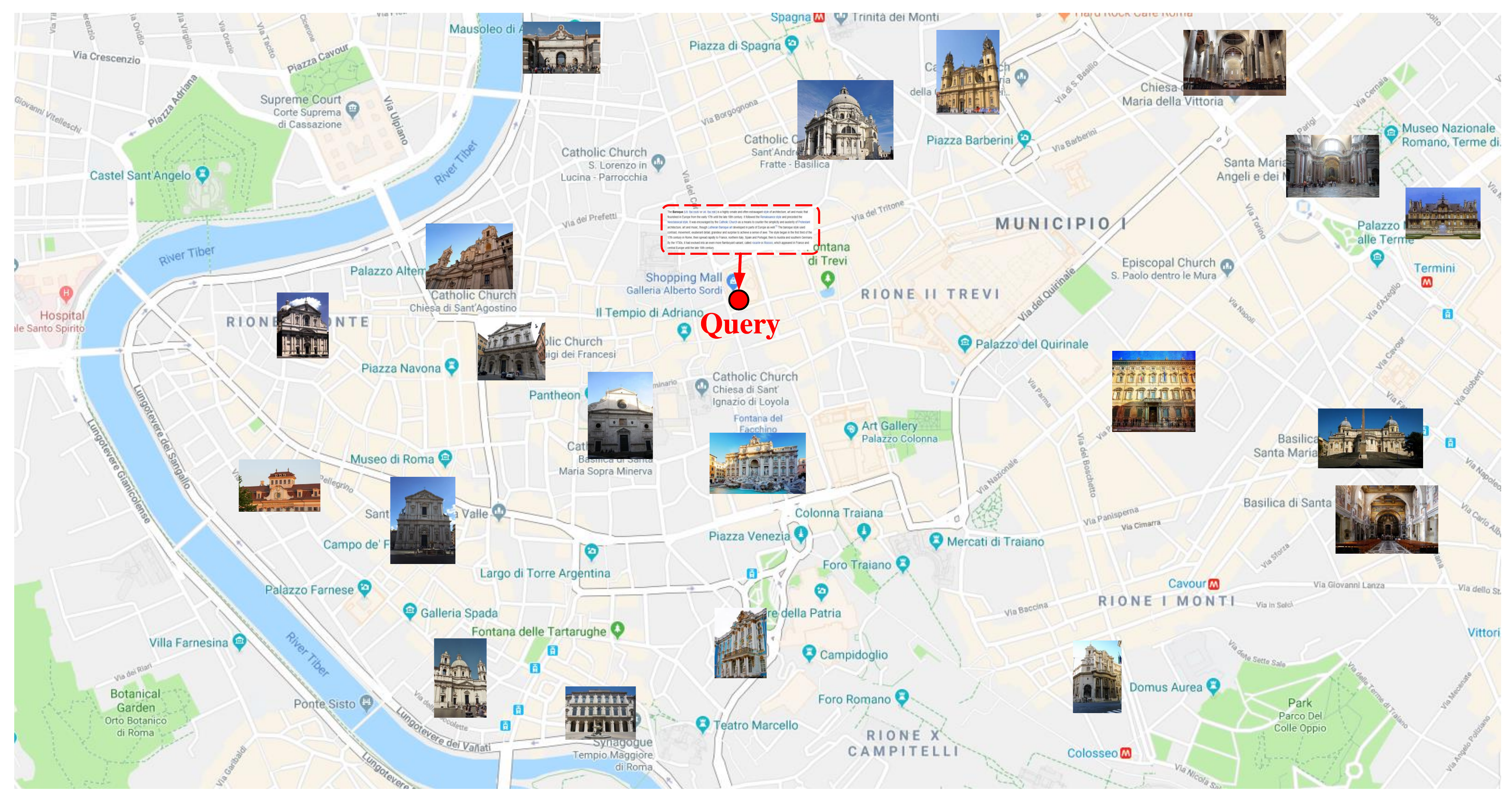}
\vspace{-1mm}
\caption{\small  An example of $k$NN spatial cross-modal retrieval }
\label{fig:fig2}
\end{figure}

\begin{example}
\label{ex:example2}
As illustrated in Figure.~\ref{fig:fig2}, consider a tourist is traveling in a historic city that has never been before. She is particularly interested in baroque architecture and She want to visit and take some photos about this type of ancient buildings. However, She have no idea how many ancient buildings are near her and she do not know where these buildings are located. She cannot seem to go all over the city to find them due to time constraints. In such case, she can write a short paragraph or a sentence to describe the desirable style of buildings or the scenery, and put them into search engine as a $k$NN spatial cross-modal query with her current location information. The system will return the $k$ nearest ancient buildings geographical location and their photos taken by other people according to her description. This tourist can find some nearest spots which meet her preferences based on the results.
\end{example}

In this paper, we aims to address the problem described in example~\ref{ex:example2}, namely, retrieval a set of results containing $k$ geo-multimedia objects which are nearest to the query location and highly similar to the query in the aspect of semantics. Based on the notion of cross-modal retrieval and spatial keyword search, we present the definition of a new type of query which is called $k$NN geo-multimedia cross-modal query and propose a novel score function which consider the geographical distance proximity and semantic similarity between two different modalities. Besides, we propose the conception of cross-modal semantic representation space and discuss the basic idea of solving cross-modal retrieval. A novel method called cross-modal semantic matching is presented which aims to construct a common semantic representation space for different modalities to bridge the semantic gap in the processing of retrieval. In order to improve the precision of search, A novel framework named DeCoSReS is designed which combines deep learning techniques (e.g., CNN) and cross-modal semantic matching. In addition, to provide the high search performance, we propose a novel hybrid indexing structure named GMR-Tree which is a combination of geo-multimedia data and R-Tree. And based on it we develop a efficient search algorithm for geo-multimedia data named $k$GMCMS to implement $k$NN geo-multimedia cross-modal query.

\noindent\textbf{Contributions.}
The main contributions of this paper can be summarized as follows:
\begin{itemize}
\item To the best of our knowledge, we are the first to propose the problem of geo-multimedia cross-modal query. We propose the definition of geo-multimedia object and $k$NN geo-multimedia cross-modal query, and then propose the conception of cross-modal semantic representation space.
\item To bridge the semantic gap between different modalities in the processing of retrieval, we propose a novel approach named cross-modal semantic matching, which consists of two important components i.e., CorrProj and LogsTran. Based on it, we design a method called DeCoSReS which uses deep learning techniques to construct cross-modal semantic representation space.
\item To improve the search performance, we present a novel hybrid indexing structure named GMR-Tree which is a combination of geo-multimedia data and R-Tree. Based on it we develop a novel efficient search algorithm named $k$GMCMS.
\item We have conducted extensive experiments on real and synthetic dataset. Experimental results demonstrate that our solution outperforms the-state-of-the-art methods.
\end{itemize}

\noindent\textbf{Roadmap.} The remainder of this paper is organized as follows: the related works are reviewed in Section 2. In Section 3 we introduce the definition of $k$NN geo-multimedia cross-modal query and related conceptions. In section 4, we propose the method named cross-modal semantic matching and then a framework of cross-modal semantic representation construction by using deep learning techniques. In Section 5, we develop a novel hybrid indexing structure named GMR-Tree and a efficient search algorithm called $k$GMCMS to support geo-multimedia cross-modal query. Our experimental results are presented in Section 6, and finally we draw our conclusion of this paper in Section 7.

\section{Related Work}
\label{related}
In this section, we introduce an overview of previous works of multi-modal and cross-modal retrieval,  multimedia retrieval based on deep learning and spatial textual search, which are related to this work. To the best of our knowledge, there is no existing work on the problem of geo-multimedia cross-modal retrieval based on deep learning technique.

\subsection{Multi-Modal and Cross-Modal Retrieval}
Multi-modal and cross-modal retrieval are two hot issues in the field of multimedia analysis and retrieval. A research problem or data set is characterized as multi-modal when it includes multiple modalities~\cite{DBLP:journals/corr/BaltrusaitisAM17} such as text, image, audio, video and 3D.  In the past few years, lots of researchers focus on multi-modal and cross-modal retrieval problem and many significant results have been proposed to improve the performance of multimedia retrieval system.

\textbf{Multi-Modal Retrieval}. Multi-modal retrieval~\cite{DBLP:journals/ivc/WuW17} method aims to search multimedia data~\cite{DBLP:conf/mm/WuWS13} with multiple modalities. Laenen et al.~\cite{DBLP:conf/wsdm/LaenenZM18} proposed a novel multi-modal fashion search paradigm, which allows users to input a multi-modal query composed of both an image and text. To address this problem, they presented a common, multi-modal space for visual and textual fashion attributes where their inner product measures their semantic similarity. Besides, they proposed a multi-modal retrieval model for fashion items search. For image raking problem, Yu et al.~\cite{DBLP:journals/tcyb/YuYGT17} proposed a novel deep multi-modal distance metric learning method named Deep-MDML to address the two main limitations of similarity estimation in existing CBIR methods: (i)Mahalanobis distance is applied to build a linear distance metric; (ii)theser methods are unsuitable for handling multi-modal data~\cite{DBLP:journals/tcyb/LiWZL17,DBLP:journals/corr/abs-1804-11013,LINYANGARXIV}. They utilized a group of autoencoders to obtain initially a distance metric in different visual spaces. Jin et al.~\cite{DBLP:journals/tip/JinLHQT18} presented a novel multi-modal hashing method named SNGH which aims to preserve the fine-grained similarity metric based on the semantic graph. They define a function based on the local similarity in particular to adaptively calculate multi-level similarity by encoding the intra-class and inter-class variations. Rafailidis et al.~\cite{DBLP:journals/pr/RafailidisMD13} designed a unified framework for multi-modal content retrieval which supports retrieval for rich media objects as unified sets of different modalities, such as image, audio, video, text and 3D. The main idea is combining all monomodal heterogeneous similarities to a global one according to an automatic weighting scheme to construct a multi-modal space to capture the semantic correlations among multiple modalities. Moon et al.~\cite{DBLP:journals/corr/MoonKW14} proposed a transfer deep learning (TDL) framework that can transfer the knowledge obtained from a single-modal neural network to a network with a different modality. Several embedding approaches for transferring knowledge between the target and source modalities were proposed by them. Dang-Nguyen et al.~\cite{DBLP:journals/tomccap/Dang-NguyenPGBN17} proposed a novel framework that can produce a visual description of a tourist attraction by choosing the most diverse pictures from community-contributed datasets to describe the queried location more comprehensively. This approach can filter out non-relevant images and to obtain a reliable set of diverse and relevant images by first clustering similar images according to their textual descriptions and their visual content. Based on multi-graph enabled active learning, Wang et al.~\cite{DBLP:conf/mir/WangMZL05} presented a multi-modal Web image retrieval technique to leverage the heterogeneous data on the Web to improve retrieval precision. In this solution, three graphes, i.e., Content-Graph, Text-Graph and Link-Graph which are constructed on visual content features, textual annotations and hyperlinks respectively, provide complimentary information on the images. In order to solve the problem of recipe-oriented image-ingredient correlation learning, Min et al.~\cite{DBLP:journals/tmm/MinJSWLH17} proposed a multi-modal multitask deep belief network (M$^3$TDBN) to learn joint image-ingredient representation regularized by different attributes.

\textbf{Cross-Modal Retrieval}. Unlike unimodal retrieval, generally the modalities of query and results are different in cross-modal retrieval, such as the retrieval of text documents in response to a query image, and the retrieval of images in response to a query text~\cite{DBLP:conf/mm/RasiwasiaPCDLLV10}. The correlation between different modalities is an important problem. In order to exploit the correlation between multiple modalities, Bredin et al.~\cite{DBLP:conf/icassp/BredinC07} utilized canonical correlation analysis (CCA) and Co-Inertia Analysis (CoIA) to solve the problem of audio-visual based talking-face biometric verification. Since the negative correlation is very important and no existing works focus on it, Zhai et al.~\cite{DBLP:conf/icassp/ZhaiPX12} proposed a novel cross-modality correlation propagation approach to simultaneously deal with positive correlation and negative correlation between media objects of different modalities. Rasiwasia et al.~\cite{DBLP:conf/aistats/RasiwasiaMMA14} proposed a novel method named cluster canonical correlation analysis (cluster-CCA) for joint dimensionality reduction of two sets of data points. Based on it they designed a kernel extension named kernel cluster canonical correlation analysis (cluster-KCCA) which achieves superior state of the art performance in cross-modal retrieval task. In another work Rasiwasia et al.~\cite{DBLP:conf/mm/RasiwasiaPCDLLV10} studied the problem of joint modeling the text and image components of multimedia documents. They investigated two hypotheses and using canonical correlation analysis to learning the correlations between text modality and image modality. To measure the cross-modal similarities, Jia et al.~\cite{DBLP:conf/iccv/JiaSD11} presented a novel probabilistic model which learns cross-modality similarity from a document corpus that has multinomial data. Based on Markov random field, this model learns a set of shared topics across the modalities. Chu et al.\cite{DBLP:journals/tcsv/ChuZLWZH16} developed a flexible multimodality graph (MMG) fusion framework to fuse the complex multi-modal data from different media and a topic recovery approach to effectively detect topics from cross-media data.

It was unfortunate that all the researches aforementioned cannot be applied to geo-multimedia cross-modal retrieval because they do not consider both the geographical location and multimedia information during a processing of multi-modal or cross-modal retrieval. These solutions are really significant for multimedia information retrieval but they are not adequately suitable to the problem of geo-multimedia cross-modal retrieval. As the multimedia data collected nowadays are always with geo-tags geographical information, which can be used to create novel multimedia search methods for advanced location-based services. Thus, there is an urgent need to develop novel approaches for geo-multimedia cross-modal retrieval.

\subsection{Multimedia Retrieval via Deep Learning}
With the rapid development of deep learning~\cite{DBLP:journals/nature/LeCunBH15}, many multimedia retrieval problems have been solve by new models via deep neural networks~\cite{DBLP:conf/nips/FromeCSBDRM13,DBLP:conf/icml/NgiamKKNLN11,DBLP:journals/cviu/WuWGHL18,DBLP:journals/pr/WuWLG18}. Content-based image retrieval is a significant problem in the area of multimedia retrieval. Recently lots of researches improve the precision of multimedia retrieval with the power of deep learning. Fu et al.~\cite{Fu2017Content} proposed a CBIR system based on CNN and SVM. In this framework, CNN is applied to extract the feature representations and SVM is used to learn the similarity measures. A validation set is generated in the training of SVM to tune to parameters. By extending SIFT-based SMK~\cite{DBLP:conf/iccv/ToliasAJ13,DBLP:journals/ijcv/ToliasAJ16a}methods, Zhou et al.~\cite{DBLP:conf/icip/ZhouLZ16} proposed a unified framework of CNN-based match kernels to encode the two complementary features: low level features and high level features, which can provide complementary information for image retrieval task. They designed a novel thresholded exponential match kernel to calculate image semantic similarity. In order to evaluate if deep learning is a hope for bridging the semantic gap in CBIR and hoe much empirical improvements can be achieved for learning feature representations and similarity measures, Wan. et al.~\cite{DBLP:conf/mm/WanWHWZZL14} investigate a framework of deep learning with application to CBIR tasks with an extensive set of empirical studies by examining a state-of-the-art deep convolutional neural network for CBIR tasks under varied settings. Sun et al.~\cite{Sun2016Learning} proposed a CNN-based image retrieval approach using Siamese network to learn a CNN model for image feature extraction. They used a contrastive loss function to enhance the discriminability of output features. Zagoruyko et al.~\cite{DBLP:conf/cvpr/ZagoruykoK15} proposed a general similarity function for patches based on CNN model to solve the problem of learning directly from raw image pixels.

\subsection{Spatial Textual Search}
Spatial textual search has been well studied for several years since this technique is significant to local-based services and advanced search engines. It aims to efficiently retrieve a set of spatial textual objects which have a high textual similarity to query keywords and are close enough to query location. Existing literatures show that there are several types of spatial textual search, such as top-$k$ search, $k$-nearest-neighbor query, range search query, etc.

A wide range of works have been conducted focus on spatial textual search and many solutions have been proposed to improve the performance of systems. R-Tree is one of the most significant spatial indexing techniques proposed by Guttman~\cite{DBLP:conf/sigmod/Guttman84}. Cao et al.~\cite{DBLP:conf/sigmod/CaoCJO11} proposed the definition of the problem of retrieving a group of spatial web objects such that the keywords of group cover the keywords of query and such that objects are nearest to the query location and have the lowest inter-object distances. Besides they proved that the two variants of this problem are NP-complete. For location-aware top-$k$ text retrieval, Cong et al.~\cite{DBLP:journals/pvldb/CongJW09} presented a new indexing framework which integrates the inverted file for text retrieval and the R-tree for spatial proximity querying. Li et al.~\cite{DBLP:conf/cikm/LiXF12} propose a new indexing framework named BR-tree by integrating a spatial component and a textual component to solve the problem of keyword-based k-nearest neighbor search in spatial databases. Zhang et al.~\cite{DBLP:conf/edbt/ZhangTT13} proposed a scalable integrated inverted index named I$^3$ based on Quadtree. Furthermore, they proposed a novel storage mechanism to improve the efficiency of retrieval and preserve summary information to pruning. In order to improve the performance of top-$k$ spatial keyword queries, Jo{\~{a}}o B. Rocha{-}Junior et al.~\cite{DBLP:conf/ssd/RochaGJN11} designed a novel index named spatial inverted index (S2I) which maps each distinct term to a set of objects containing the term. Li et al.~\cite{DBLP:journals/tkde/LiLZLLW11} introduced a novel index named IR-Tree which indexes both the textual and spatial contents of documents to support document retrieval and then designed a top-$k$ document search algorithm. Zhang et al.~\cite{DBLP:conf/sigir/ZhangCT14} proposed an effective approach to solve the top-$k$ distance-sensitive spatial keyword query by modeling it as the well-known top-$k$ aggregation problem. Besides, they developed a novel algorithm called Rank-aware CA (RCA) algorithm based on CA algorithm. Zhang et al.~\cite{DBLP:conf/icde/ZhangCMTK09} introduced a novel novel spatial keyword query problem named $m$-closest keywords ($m$CK) query which aims to search out the spatially closest tuples which match $m$ user-specified keywords. To solve this problem more efficiently, they designed a novel index called the bR$^*$-tree extended from R$^*$-tree~\cite{DBLP:conf/sigmod/BeckmannKSS90}. Moreover, They exploited a priori-based search strategies to effectively reduce the search space. For collective spatial keyword query problem, Long et al.~\cite{DBLP:conf/sigmod/LongWWF13} proposed a distance owner-driven method including an exact algorithm that runs faster than the best-known existing algorithm and an approximate algorithm which improves the constant approximation factor from 2 to 1.375. For top-$k$ spatial keyword search problem, Zhang et al.~\cite{DBLP:conf/icde/ZhangZZL13} presented a novel index structure named inverted linear quadtree (IL-Quadtree) based on inverted index and the linear quadtree. Furthermore, they developed an efficient algorithm to improve the efficiency of search.

It is apparently that these solutions for spatial textual search problem aforementioned just only consider the situation that the geo-location objects which contain only one modality, i.e., text or keywords. In other words, These methods cannot be applied to spatial cross-modal retrieval in the geo-multimedia database. This necessitates the development of novel and efficient cross-modal search methods for multimedia data with geographical information. To the best of our knowledge, we are the first to study the problem of geo-multimedia cross-modal retrieval considering both different features of multimodality data and the geographical information which is extracted from the geo-tags of geo-multimedia data.

\section{Preliminary}
\label{preliminary}

In this section, we firstly formally define the geo-multimedia object and some relevant concepts, then we introduce the definition of $k$NN geo-multimedia cross-modal query. Furthermore, we propose the conception of cross-modal semantic representation mapping. Table ~\ref{tab:notation} summarizes the mathematical notations used throughout this paper to facilitate the discussion of our work.

\begin{table}
	\centering
    \small
	\begin{tabular}{|p{0.14\columnwidth}| p{0.78\columnwidth} |}
		\hline
		\textbf{Notation} & \textbf{Definition} \\ \hline\hline
		~$\mathcal{O}$                                   & A given database of geo-multimedia objects      \\ \hline
        ~$|\mathcal{O}|$                                 & The number of objects in $\mathcal{O}$     \\ \hline
        ~$o.\lambda$                                     & The geo-location information descriptor of $o$    \\ \hline
	  	~$o.\psi$                                        & A visual content descriptor of $o$        \\ \hline
        ~$\mathcal{Q}^k$                                 & A $k$NN geo-multimedia cross-modal query      \\ \hline
        ~$\mathcal{Q}_{T2I}$                             & A text query to search images      \\ \hline
        ~$\mathcal{Q}_{I2T}$                             & A image query to search texts      \\ \hline
        ~$\mathcal{M}$                                   & A modality set      \\ \hline
		~$T$                                             & Text modality    \\ \hline
        ~$I$                                             & Image modality    \\ \hline
		~$\mathbb{S}_T$                                  & A text feature space             \\ \hline
        ~$\mathbb{S}_I$                                  & A image feature space             \\ \hline
        ~$M_T$                                           & a feature vector of a text             \\ \hline
        ~$M_I$                                           & a feature vector of an image             \\ \hline
        ~$X$                                             & The longitude of a geo-location              \\ \hline
        ~$Y$                                             & The latitude of a geo-location              \\ \hline
        ~$k$                                             & The number of final results              \\ \hline
        ~$\mathcal{F}_{score}(\mathcal{Q},o)$            & Score function measuring the similarity of $\mathcal{Q}$ and $o$\\ \hline
        ~$\mathcal{R}$                                   & The set of results      \\ \hline
        ~$\mu$                                           & A parameter to balance distance proximity and semantic similarity  \\ \hline
        ~$Dst(\mathcal{Q},o)$                            & The spatial distance function.    \\ \hline
        ~$\delta(\mathcal{Q},o)$                         & The Euclidean distance between $\mathcal{Q}$ and $o$\\ \hline	
    	~$Sim(\mathcal{Q},o)$                            & The semantic similarity between $\mathcal{Q}$ and $o$\\ \hline
        ~$\mathbb{W}_T$                                  & An intermediate representation space of text modality  \\ \hline
        ~$\mathbb{W}_I$                                  & An intermediate representation space of image modality  \\ \hline
        ~$\mathbb{R}_T$                                  & The semantic representation space of text modality \\ \hline
        ~$\mathbb{R}_I$                                  & The semantic representation space of image modality \\ \hline
        ~$\Psi$                                          & A mapping from text feature space to image feature space\\ \hline
        ~$\mathfrak{W}$                                  & A cross-modal semantic representation space \\ \hline
        ~$\mathfrak{L}_T$                                & A non-linear transformation for text modality \\ \hline
        ~$\mathfrak{L}_I$                                & A non-linear transformation for image modality \\ \hline
        ~$\Theta_T$                                      & A projection from text feature space to intermediate representation space  \\ \hline
        ~$\Theta_I$                                      & A projection from image feature space to intermediate representation space  \\ \hline
        ~$\mathcal{C}$                                   & The set of semantic concepts   \\ \hline
        ~$\Upsilon$                                      & The set of classes \\ \hline
        ~$N_i$                                           & A node of GMR-Tree \\ \hline
        ~$S_i$                                           & A signature \\ \hline
        ~$\bigwedge$                                     & The  binary OR-ing operation \\ \hline
        ~$\mathcal{H}_{SIG}(.)$                          & A hashing function to generate signatures \\ \hline
	\end{tabular}
    \caption{The summary of notations} \label{tab:notation}	
\end{table}

\subsection{Problem Definition}
\begin{definition}[\textbf{Geo-Multimedia Object}] \label{def:igis gmo}
A geo-multimedia objects database is defined as $\mathcal{O}=\{o_1, o_2,..., o_{|\mathcal{O}|}\}$, in which $|\mathcal{O}|$ represents the number of objects in $\mathcal{O}$. Each geo-multimedia object $o \in \mathcal{O}$ is associated with a geographical information descriptor $o.\lambda$ and a modality content descriptor $o.M$. a geographical information descriptor includes a 2-dimensional geographical location with longitude $X$ and latitude $Y$ is represented by $o.\lambda = (X,Y)$. Let $\mathcal{M}$ be the modality set. In this paper we just consider two most common modalities, i.e., text and image, thus $\mathcal{M} = \{T,I\}$, where $T$ represents text modality and $I$ represents image modality. If a geo-multimedia object contains a text, it is denoted as $o.M_{T}$. Similarly, If an object contains an image, it is denoted as $o.M_{I}$. $M_T$ and $M_I$ denote the feature vector generated by a text and an image respectively. Let $\mathbb{S}_{T}$ and $\mathbb{S}_{I}$ be the feature spaces of text and image, $\forall o_i \in \mathcal{O}$, if $o_i$ contains a text, then $o_i.M_T \in \mathbb{S}_T$. If $o_i$ contains an image, then $o_i.M_I \in \mathbb{S}_I$.
\end{definition}

Based on the definition of geo-multimedia objects, we can define the $k$NN geo-multimedia cross-modal query. Firstly we consider the query without geographical information. In other words, we define the cross-modal query and then extend it to the query in the geo-multimedia database.

\begin{definition}[\textbf{Coss-Modal Query}] \label{def:igis gmcmq}
Given a multimedia objects database $\mathcal{O}=\{o_1, o_2,..., o_{|\mathcal{O}|}\}$, in which each object contains one of the following two modalities, i.e., text modality $T$ and image modality $I$. There are two types of cross-modal query can be defined: (1) $\mathcal{Q}_{T2I}$ is defined as a text query which aims to search our the most relevant multimedia object $o \in \mathcal{O}$ contains an image, and $\mathcal{Q}_{T2I}.M_T \in \mathbb{S}_T$,$o.M_I \in \mathbb{S}_I$. (2)$\mathcal{Q}_{I2T}$ is defined as a image query which aims to search out the most relevant multimedia object $o \in \mathcal{O}$ contains a text, and $\mathcal{Q}_{I2T}.M_I \in \mathbb{S}_I$,$o_i.M_T \in \mathbb{S}_T$.
\end{definition}

\begin{definition}[\textbf{$k$NN Geo-Multimedia Cross-Modal Query}] \label{def:igis gmcmq}
Given a geo-multime\\dia objects database $\mathcal{O}=\{o_1, o_2,..., o_{|\mathcal{O}|}\}$, a $k$NN Geo-Multimedia Cross-Modal Query $\mathcal{Q}^k = (\lambda,M)$ aims to return $k$ nearest geo-multimedia objects whose modalities features are highly relevant to the query. Like Definition~\ref{def:igis gmcmq}, we define these two types of query as $\mathcal{Q}^k_{T2I}$ and $\mathcal{Q}^k_{I2T}$, which are named $k$NN geo-multimedia text to image query ($k$T2IQ for short) and $k$NN geo-multimedia image to text query ($k$I2TQ for short) respectively. In more detail, $\mathcal{Q}^k_{T2I}$ aims to return $k$ nearest geo-multimedia objects which contain images that are highly relevant to the query text, and $\mathcal{Q}^k_{I2T}$ aims to find $k$ nearest objects which contain texts that are highly relevant to the query image. The relevancy between text and image is the semantic correlation between them. Formally, For query $\mathcal{Q}^k_{T2I}$, the result is $k$ geo-multimedia objects $R_{T2I}$ which are ranked by the a score function $\mathcal{F}_{score}(\mathcal{Q}^k_{T2I},o)$,i.e.,
\begin{equation*}\label{knngmcmqt2i}
\mathcal{R}_{T2I} = \{o|\forall o \in \mathcal{O},o' \in \mathcal{O} \setminus \mathcal{R}_{T2I},\mathcal{F}_{score}(\mathcal{Q}^k_{T2I},o)>\mathcal{F}_{score}(\mathcal{Q}^k_{T2I},o')\},
\end{equation*}
\begin{equation*}\label{knngmcmqt2i}
\mathcal{R}_{T2I} \subseteq \mathcal{O}, |\mathcal{R}_{T2I}| = k
\end{equation*}
likewise, for query $\mathcal{Q}^k_{I2T}$, the result is $k$ geo-multimedia objects $R_{T2I}$ ranked by $\mathcal{F}_{score}(\mathcal{Q}^k_{I2T},o)$, i.e.,
\begin{equation*}\label{knngmcmqi2t}
\mathcal{R}_{I2T} = \{o|\forall o \in \mathcal{O},o' \in \mathcal{O} \setminus \mathcal{R}_{I2T},\mathcal{F}_{score}(\mathcal{Q}^k_{I2T},o)>\mathcal{F}_{score}(\mathcal{Q}^k_{I2T},o')\},
\end{equation*}
\begin{equation*}\label{knngmcmqt2i}
\mathcal{R}_{I2T} \subseteq \mathcal{O}, |\mathcal{R}_{I2T}| = k
\end{equation*}
and the score function is defined as follows:
\begin{equation}\label{equ:scorefunc}
\mathcal{F}_{score}(\mathcal{Q},o) = \mu Dst(\mathcal{Q},o)+(1-\mu)Sim(\mathcal{Q}.o)
\end{equation}
where $\mathcal{Q}$ represents a query, and $\mu \in [0,1]$ is a parameter which is to balance the importance between distance proximity component and semantic similarity component. If $\mu > 0.5$, it means the distance proximity is more important than the semantic similarity. And if $\mu = 0$, it means this function is just used to measure the semantic similarity between $\mathcal{Q}$ and $o$.
\end{definition}

In this paper, we just only concentrate on the $k$T2IQ query $\mathcal{Q}^k_{T2I}$. That is, given a query text, the system will measure the geographical distance proximity according the location information of query and objects, and meanwhile calculate the relevance between query text and images contained in objects. Thus we abbreviate $\mathcal{Q}^k_{T2I}$ as $\mathcal{Q}$. In the following part we introduce how to measure spatial distance proximity between a query and an object, and the semantic correlation between text and image.

\begin{definition}[\textbf{Spatial distance proximity measurement}] \label{def:igis sdpm}
Given a geo-multimedia objects database $\mathcal{O}=\{o_1, o_2,..., o_{|\mathcal{O}|}\}$ and a $k$T2IQ query $\mathcal{Q}$, $\forall o \in \mathcal{O}$, the spatial distance proximity is measured by the following function:
\begin{equation}\label{equ:sdpm}
Dst(\mathcal{Q},o) = 1 - \frac{\delta(\mathcal{Q},o)}{\delta_{max}(\mathcal{Q},\mathcal{O})}
\end{equation}
where $\delta(\mathcal{Q},o)$ represents Euclidean distance between the query $\mathcal{Q}$ and the object $o$. $\delta_{max}(\mathcal{Q},\mathcal{O})$ represents the maximum spatial distance between $\mathcal{Q}$ and any objects in $\mathcal{O}$. They are defined in detail as follows:
\begin{equation}\label{equ:eculideandst}
\delta(\mathcal{Q},o) = \sqrt{(\mathcal{Q}.\lambda.X-o.\lambda.X)^2+(\mathcal{Q}.\lambda.Y-o.\lambda.Y)^2}
\end{equation}
\begin{equation}\label{equ:maxdst}
\delta_{max}(\mathcal{Q},\mathcal{O}) = max(\{\delta(\mathcal{Q},o)|\forall o \in \mathcal{O}\})
\end{equation}
where the function $max(\mathcal{X})$ is to return the maximum value of element in the set $\mathcal{X}$. It is easily to know that for spatial distance proximity measurement, the objects with the \textbf{small score values} are preferred (i,e., ranked higher).
\end{definition}

\begin{definition}[\textbf{Cross-modal semantic similarity measurement}] \label{def:igis sdpm}
Given a geo-multimedia objects database $\mathcal{O}=\{o_1, o_2,..., o_{|\mathcal{O}|}\}$ and a $k$T2IQ query $\mathcal{Q}$, $\forall o \in \mathcal{O}$, the cross-modal semantic similarity is measured by cosine similarity measurement, as shown in the following equation:
\begin{equation}\label{equ:cmssm}
Sim(\mathcal{Q},o) = \frac{\sum_{i \in \mathcal{Q}.M_T}^{}\mathcal{Q}.M_T^{(i)} * o.M_I^{(i)}}{\sqrt{\sum_{i \in \mathcal{Q}.M_T}^{}(\mathcal{Q}.M_T^{(i)})^2}*\sqrt{\sum_{i \in o.M_I}^{}(o.M_I^{(i)})^2}}
\end{equation}
where $\mathcal{Q}.M_T^{(i)}$ and $o.M_I^{(i)}$ represent $i$th feature element in representation vector $\mathcal{Q}.M_T$ and $o.M_I$ respectively.
\end{definition}

\subsection{Cross-Modal Semantic Representation Space}\label{subsec:cmsrs}
It is common knowledge that semantic gaps exist between different modalities, which is a ticklish problem for cross-modal retrieval. In other words, we cannot directly calculate similarity between query and object which belongs to different modalities by equation~\eqref{equ:cmssm} because $\mathcal{Q}.M_T$ and $o.M_T$ cannot be mapped into a common space. Therefore, this task cannot be reduced to a classical information retrieval task in which there is a mapping between query representation space and object representation space. This can be described in formal as follows: for a query $\mathcal{Q}$ with a text and a geo-multimedia object $o$ with an image, the features spaces of them are denoted as $\mathbb{S}_T$ and $\mathbb{S}_I$ respectively, and $\mathcal{Q}.M_T \in \mathbb{S}_T, o.M_I \in \mathbb{S}_I$, the mapping between $\mathbb{S}_T$ and $\mathbb{S}_I$ is represented as
\begin{equation*}
\Psi: \mathbb{S}_T \longrightarrow \mathbb{S}_I
\end{equation*}
and the inverse mapping is represented as
\begin{equation*}
\Psi^{-1}: \mathbb{S}_I \longrightarrow \mathbb{S}_T
\end{equation*}
Thus, the cross-modal text to image query can be denoted as $\mathcal{Q}_{T2I} \iff \Psi(\mathcal{Q}.M_T)$. As discussed above, it is hard to find this mapping between feature spaces of different modalities.

To address this problem, we assume that there exist two mappings which map text and image feature spaces into two intermediate representation $\mathbb{W}_T$ and $\mathbb{W}_I$ respectively, that is:
\begin{equation*}
\Omega_T: \mathbb{S}_T \longrightarrow \mathbb{W}_T
\end{equation*}
\begin{equation*}
\Omega_I: \mathbb{S}_I \longrightarrow \mathbb{W}_I
\end{equation*}
and the inverse mappings of them are denoted respectively as
\begin{equation*}
\Omega^{-1}_T: \mathbb{W}_T \longrightarrow \mathbb{S}_T
\end{equation*}
\begin{equation*}
\Omega^{-1}_I: \mathbb{W}_I \longrightarrow \mathbb{S}_I
\end{equation*}
and existing a mapping $\Phi$:
\begin{equation*}
\Phi: \mathbb{W}_T \longrightarrow \mathbb{W}_I
\end{equation*}
that means there is a semantic correlation between these two isomorphic spaces $\mathbb{W}_T$ and $\mathbb{W}_I$.

Based on this assumption, we can redescribe the cross-modal text to image query in the following forms: Given a geo-multimedia database $\mathcal{O}$, a $k$T2IQ query $\mathcal{Q}$ is to search out the most relevant object contains image that is represented as $\Omega^{-1}_I(\Phi(\Omega_T(\mathcal{Q}.M_T)))$ in $\mathbb{S}_I$. In other words, This idea is to apply two intermediate representation spaces $\mathbb{W}_T$ and $\mathbb{W}_I$ to implement the mapping from $\mathbb{S}_T$ to $\mathbb{S}_I$.

According to the above discussion, the most difficult problem for implementing efficient cross-modal retrieval is to learn the intermediate representation spaces $\mathbb{W}_T$ and $\mathbb{W}_I$. To overcome this challenge, we introduce a notion named \textbf{C}r\textbf{O}ss-modal \textbf{S}emantics \textbf{Re}presentation \textbf{S}pace (CoSReS for short) and the definition is shown as follows.

\begin{figure}[thb]
\newskip\subfigtoppskip \subfigtopskip = -0.1cm
\centering
\includegraphics[width=1.0\linewidth]{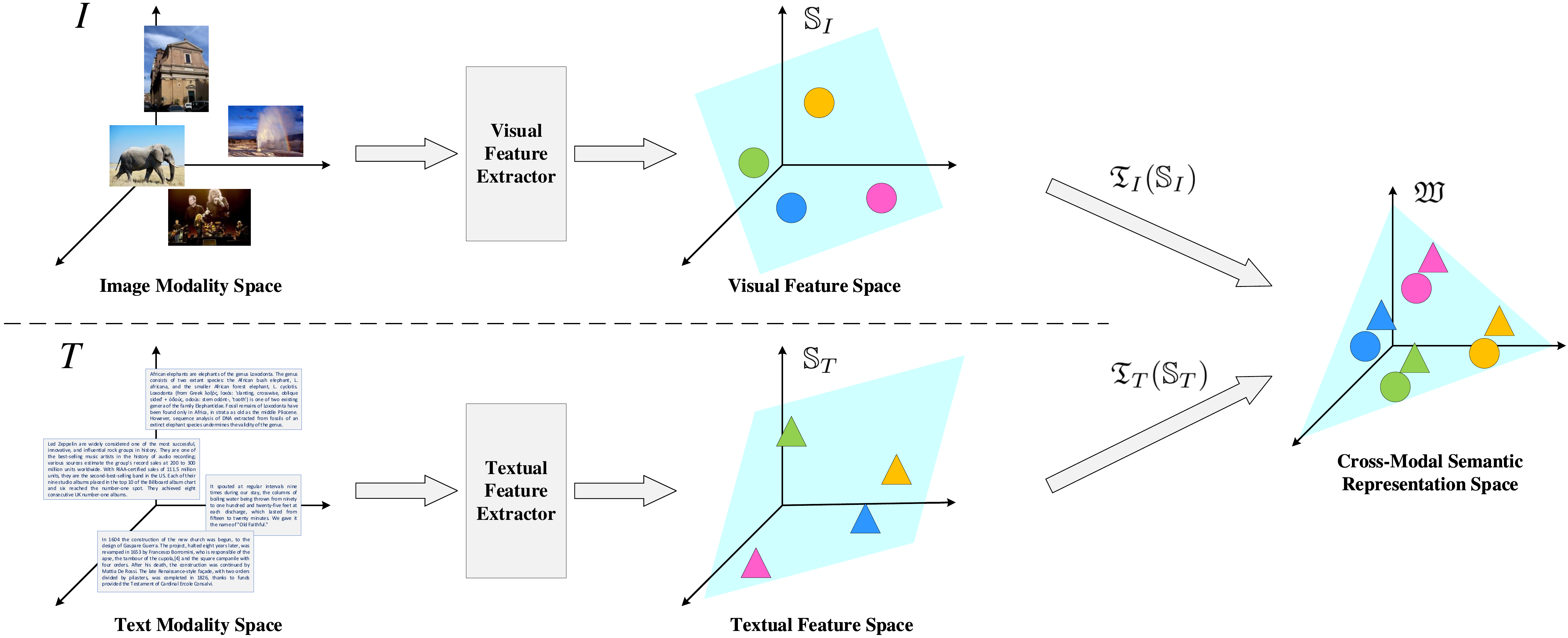}
\vspace{-1mm}
\caption{\small  Cross-modal semantics representation space }
\label{fig:fig3}
\end{figure}

\begin{definition}[\textbf{Cross-Modal Semantic Representation Space (CoSReS)}] \label{def:cosres}
Given a geo-multimedia database $\mathcal{O}$ and modality set $\mathcal{M} = \{T,I\}$. Let $\mathbb{S}_T$ and $\mathbb{S}_I$ be the feature spaces of text and image respectively, $\mathbb{R}_T$ and $\mathbb{R}_I$ be the semantic space of text and image respectively. A cross-modal semantic representation space (CoSReS) $\mathfrak{W}$ is a isomorphic representation space for modalities $T$ and $I$ in a high-level semantic abstraction, if existing two non-linear transformations $\mathfrak{T}_T$ and $\mathfrak{T}_I$, $\mathbb{R}_T = \mathfrak{T}_T(\mathbb{S}_T)$ and $\mathbb{R}_I = \mathfrak{T}_I(\mathbb{S}_I)$, then $\mathfrak{W} = \mathbb{R}_T = \mathbb{R}_I$.
\end{definition}

Figure.~\ref{fig:fig3} demonstrates the conception of CoSReS. For two different modalities, CoSReS have a set of common semantic conceptions. After extracting features for texts and images respectively, the feature vectors of texts and images can be transformed into semantic representation vectors in CoSReS. Therefore, we can easily calculate the semantic similarity in this common representation space.

\section{Cross-Modal Semantic Representation Space construction with Deep Learning}
\label{DeCoSReS}

In the last section, we present that we can reduce the task of bridging the semantic gaps between different modalities into the problem of intermediate representation space construction, which can be represented by cross-modal semantic representation space (CoSReS). In this section, we present a solution with on deep learning techniques to construct the CoSReS based on the conception presented in subsection~\ref{subsec:cmsrs}. First we discuss how to learn a common semantic representation space for text and image data. Then an effective approach named DeCoSReS is introduced, which utilizes convolution neural networks (CNN for short) and latent Dirichlet allocation~\cite{Blei2003Latent} (LDA for short) to learn the representation speace.

\subsection{Cross-Modal Semantic Matching}
We use the method called cross-modal semantic matching (CoSMat for short) to construct CoSReS so that it provides a common semantic representation space for different modalities. This algorithm consists of two components, i.e., (1)CCA based correlation projection (CorrProj for short) and (2)logistic regression based transformation (LogsTran). The former aims to learn subspaces from feature spaces of different modalities, and the latter is to learn semantic mappings in these subspaces. We introduce these two important techniques respectively in the following part.

\noindent\textbf{(1)CorrProj}. Canonical correlation analysis~\cite{Hotelling1936Relations} (CCA) is a popular dimensionality reduction method. We use it to learn $\gamma$-dimensional subspace $\mathbb{W}_T^\gamma \in \mathbb{S}_T$ and $\mathbb{W}_I^\gamma \in \mathbb{S}_I$ to find the correlations between these two subspaces. CCA method learns directions in text and image feature spaces, i.e., $\Gamma_T \in \mathbb{S}_T$ and $\Gamma_I \in \mathbb{S}_I$ along the directions of the data maximally correlated. That is, for feature vectors $M_T$ and $M_I$, calculating the maximun correlation:
\begin{equation*}
\mathfrak{u} = \Gamma_T^TM_T,
\end{equation*}
\begin{equation*}
\mathfrak{v} = \Gamma_I^TM_I,
\end{equation*}
\begin{equation}\label{equ:maxcorr}
max Corr(\mathfrak{u},\mathfrak{v}) = \frac{\Gamma_T^T \Sigma_{TI}\Gamma_I}{\sqrt{\Gamma_T^T \Sigma_{TT}\Gamma_T}\sqrt{\Gamma_I^T \Sigma_{II}\Gamma_I}}
\end{equation}
wherein $\Sigma_{TT}$ and $\Sigma_{II}$ are the empirical covariance matrices of space $\mathbb{S}_T$ and $\mathbb{S}_I$, i.e., $\Sigma_{TT} = Cov(\mathbb{S}_T)$ and $\Sigma_{II} = Cov(\mathbb{S}_I)$, $\Sigma_{TI}$ is the empirical cross-covariance matrix of them, i.e., $\Sigma_{TI} = XCov(\mathbb{S}_T,\mathbb{S}_I)$, and $\Sigma_{TI} = \Sigma_{IT}^T$.

The first $\gamma$ canonical components $\{\Gamma_{T_1}\}^\gamma$ and $\{\Gamma_{I_1}\}^\gamma$ represent a basis for projection $\mathbb{S}_T$ and $\mathbb{S}_I$ on subspace $\mathbb{W}_T$ and $\mathbb{W}_I$. For each text $M_T$ in space $\mathbb{S}_T$, it can be mapped into the projection $\Theta_T(M_T)$ onto $\{\Gamma_{T_1}\}^\gamma$. Likewise, for each image $M_I$ in space $\mathbb{S}_I$, it can be mapped into the projection $\Theta_I(M_I)$ onto $\{\Gamma_{I_1}\}^\gamma$. Therefore, the method CorrProj can learn two projections $\Theta_T(M_T)$ and $\Theta_I(M_I)$ from $\mathbb{S}_T$ and $\mathbb{S}_I$, which can be used to define two $\gamma$-dimension subspaces for text and image, i.e.,
\begin{equation*}
\Theta_T: \mathbb{S}_T \longrightarrow \mathbb{W}_T
\end{equation*}
and,
\begin{equation*}
\Theta_I: \mathbb{S}_I \longrightarrow \mathbb{W}_I
\end{equation*}

After that, this approach used another component named LogsTran to learn two semantic mappings from these two subspace, which is described as follows.

\noindent\textbf{(2)LogsTran}. The method aforementioned is to map feature spaces of text and image to maximally correlated subspaces $\mathbb{W}_T$ and $\mathbb{W}_I$. Then we use another method called LogsTran to find the correspondence between $\mathbb{S}_T$ and $\mathbb{S}_I$ by represented objects at a higher-level of semantic abstraction. It can map text and image space into a common semantic representation space with a set of semantic concepts $\mathcal{C} = \{c_1,c_2,...,c_n\}$, such as "airplane","cat" or "house". We utilize logistic regression to learn two transformation $\mathfrak{L}_T$ and $\mathfrak{L}_I$. $\mathfrak{L}_T$ transforms a text contained by a geo-multimedia object $o.M_T \in \mathbb{S}_T$ into a vector of posterior probabilities $P_{T}^\Upsilon(\upsilon_i|T)$, in which $\Upsilon = \{\upsilon_1,\upsilon_2,...,\upsilon_k\}$ is a set of classes. Likewise, $\mathfrak{L}_I$ transforms an image contained by a geo-multimedia object $o.M_I \in \mathbb{S}_I$ into a vector of posterior probabilities $P_{I}^\Upsilon(\upsilon_i|I)$. The spaces $\mathbb{R}_T$ and $\mathbb{R}_I$ of these posterior probabilities vectors are referred to the semantic representation space of text and image respectively. Formally, they can be presented as follows:
\begin{equation*}
\mathfrak{L}_T: \mathbb{S}_T \longrightarrow \mathbb{R}_T
\end{equation*}
\begin{equation*}
\mathfrak{L}_I: \mathbb{S}_I \longrightarrow \mathbb{R}_I
\end{equation*}

Multi-calss logistic regression is utilized, which produces a linear classifier. It calculates the posterior probability of class $c_i$ by the following logistic function:
\begin{equation}\label{equ:logisticfunc}
P_M^\Upsilon(c_i|M_x;\varpi) = \frac{1}{\sum_{c_i}^{}exp(\varpi_{c_i}^TM_x)}exp(\varpi_{c_i}^TM_x)
\end{equation}
where $M$ represents the modalities information. For example, for text, $M = T$ and for image , $M = I$. $M_x$ is the features vector in the input space. $\varpi = (\varpi_1,\varpi_2,...,\varpi_k)$ is a vector of parameters for class $c_i$.

According to the logistic regression, in semantic representation spaces $\mathbb{R}_T$ and $\mathbb{R}_I$, the features are semantic conception probabilities, for instance, the probability of a text belongs to "cat" class or the probability of an image belongs to "airplane" class. Furthermore, texts and images are represented as posterior probabilities vectors in regard to same classes. In addition, the semantic representation spaces $\mathbb{R}_T$ and $\mathbb{R}_I$ are isomorphic, and they can be regarded as the same, that is, $\mathbb{R}_T = \mathbb{R}_I$. Therefore, the cross-modal semantic representation space $\mathfrak{W} = \mathbb{R}_T = \mathbb{R}_I$.

The CosMat method is a combination of CorrProj and LogsTran. In the first step, CorrProj is applied to learn two maximally correlated subspaces $\mathbb{W}_T$ and $\mathbb{W}_I$ based on feature spaces $\mathbb{S}_T$ and $\mathbb{S}_I$. Then LogsTran method is used to generate two transformations $\mathfrak{L}_T$ and $\mathfrak{L}_I$ to create the isomorphic semantic representation spaces $\mathbb{R}_T$ and $\mathbb{R}_I$. Thus, we can measure the semantic similarity of text and image in the CoSReS $\mathfrak{W}$, i.e., $Sim(\xi_T,\xi_I)$, where $\xi_T = \mathfrak{L}_T(\Theta_T(\mathbb{S}_T))$, $\xi_I = \mathfrak{L}_I(\Theta_I(\mathbb{S}_I))$. It is an significant step of implementing $k$T2IQ.

\subsection{Cross-Modal Semantic Representation Space Learning}
\begin{figure}[thb]
\newskip\subfigtoppskip \subfigtopskip = -0.1cm
\centering
\includegraphics[width=1.0\linewidth]{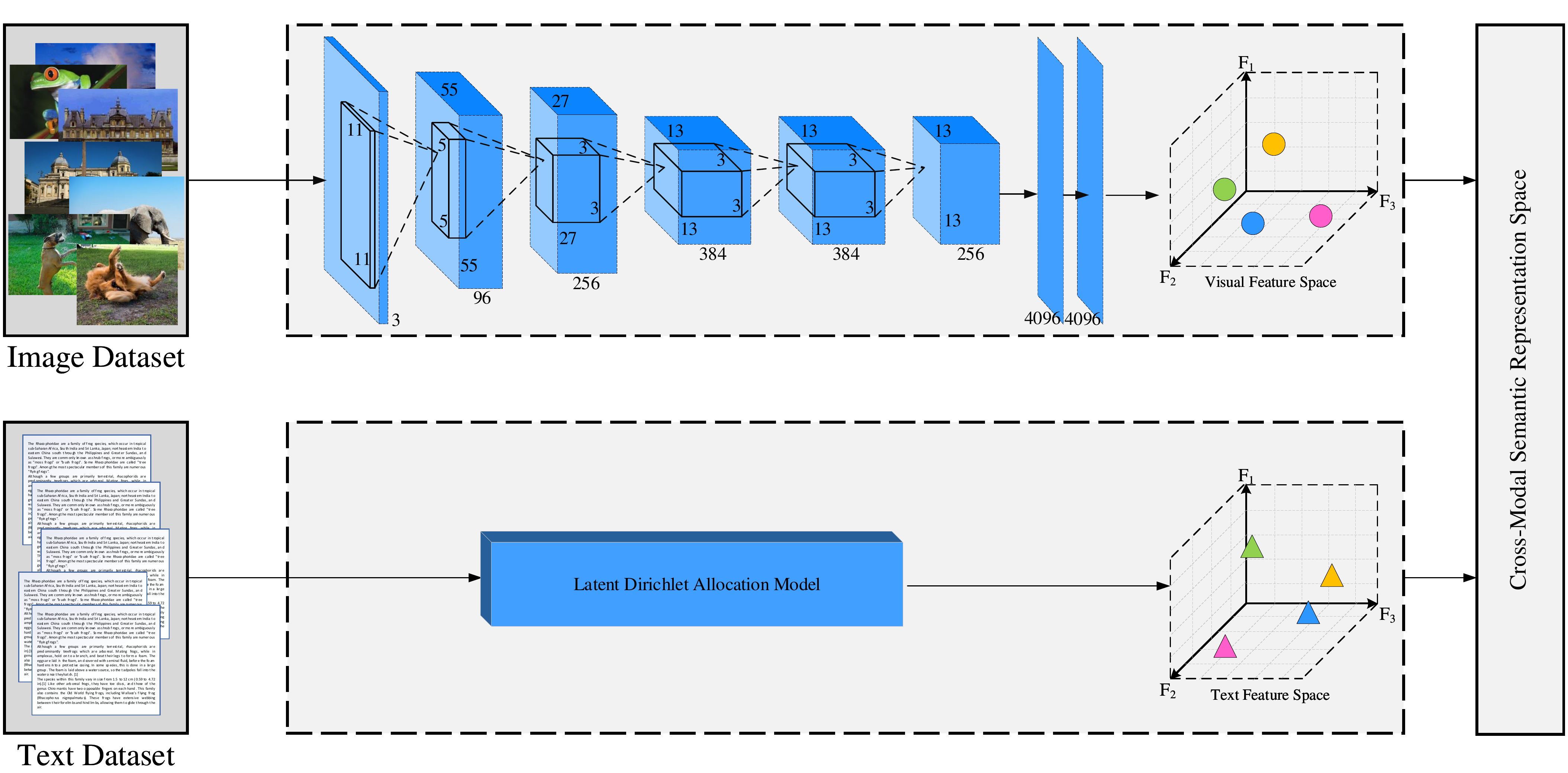}
\vspace{-1mm}
\caption{\small  The framework of cross-modal semantic representation space construction with deep learning }
\label{fig:fig4}
\end{figure} 

Deep learning techniques such as CNN, RNN, etc. are widely applied in the area of multimedia retrieval. To implement cross-modal semantic representation space construction and cross-modal retrieval, we utilize CNN to extract visual features from images and use LDA model and fully-connected networks to extract textual features. Figure~\ref{fig:fig4} is the framework of cross-modal semantic representation space construction with deep learning.

For visual features extraction, The CNN model used in this framework contains five convolutional layers and two fully-connected layers, which is trained by 1 million images. Specifically, each image is first resized to $256 \times 256$ and input this CNN model. The fully-connected layers denote 4096 dimensional features after ReLU. 

For textual feature extraction, we utilize latent Dirichlet allocation (LDA) model to generate the representation of the input text. LDA is a generative model for a text corpus in which the semantic content of a text is summarized as a mixture of serval topics. Specifically, a text is modeled by a multinomial distribution over $\kappa$ topics and each word in a text is generated by first sampling a topic from the text-speccific topic distribution~\cite{Blei2003Latent}. 
\section{Hybird Indexing for Geo-Multimedia Cross-Modal Retrieval}
\label{HybirdIndexing}

In this section, we present a novel hybrid spatial indexing technique for efficient geo-multimedia cross-modal retrieval, named \textbf{G}eo-\textbf{M}ultimedia \textbf{R-Tree} (GMR-Tree for short). Firstly we introduce the basic structure of GMR-Tree and related concepts. Then we propose our search algorithm which can boost the performance of geo-multimedia cross-modal query.

\subsection{Hybrid Indexing Structure}
The novel hybrid indexing structure proposed in this paper is called GMR-Tree which is a combination of an R-Tree~\cite{DBLP:conf/sigmod/Guttman84} and signature files. Different from R-Tree, the nodes of GMR-Tree not only contain geo-location information, but carry modality semantic representation information as well. The geo-location information is represented in the form of minimum bounding area and semantic representation information is in the form of a signature. In the following part, we introduce this novel indexing technique in more detail.

\begin{figure}[thb]
\newskip\subfigtoppskip \subfigtopskip = -0.1cm
\centering
\includegraphics[width=1.0\linewidth]{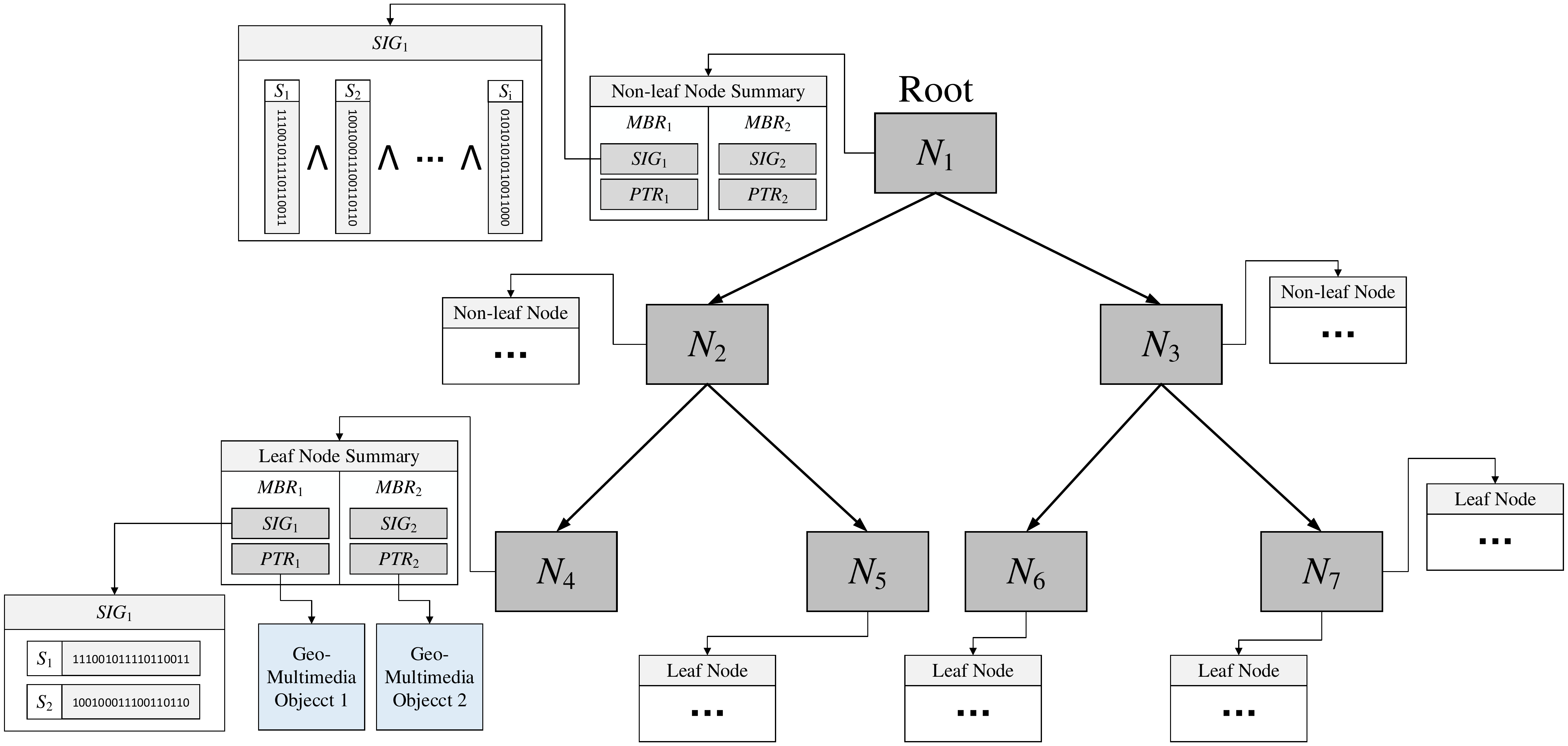}
\vspace{-1mm}
\caption{\small  A GMR-Tree }
\label{fig:fig5}
\end{figure}

Figure.~\ref{fig:fig5} illustrates the basic indexing structure of a GMR-Tree. Generally, a GMR-Tree is a height-balanced tree structure. Each non-leaf node denoted as a triple $<MBR,SIG,PTR_N>$ contains three components. $MBR$ is defined as in the R-Tree, which represents the geo-location in the form of a minimum bounding area (MBR for short). $SIG$ is a signature file generated from the geo-multimedia objects in this $MBR$. For the $i$th object in $MBR$ $o_i$, its signature is denoted as $S_i = \mathcal{H}_{SIG}(o_i.M_I)$, where $\mathcal{H}_{SIG}(.)$ is a hashing function which is used to generate a signature from a semantic representation vector of a object. For a $MBR_1$, the signature $SIG_1 = S_1 \bigwedge S_2 \bigwedge ... \bigwedge S_i$, wherein the operator $\bigwedge$ represents binary OR-ing operation. In other words, the signature of a node is equivalent to a signature that superimposes the signatures of the children nodes. In addition, the length of the signatures in each level is the same. The third component of node is a pointer $PTR_N$, which is used to point to a subnode. Similarly, the leaf note in GMR-Tree is the form of $<MBR,SIG,PTR_o>$ but the pointer $PTR_o$ is used to point geo-multimedia objects, rather than subnode.

According to the structure of GMR-Tree and the calculation of signature, we can find a very useful property of GMR-Tree which can provide well support for the spatial search. We describe it in \emph{Property}~\ref{prop:gmr-tree}.

\begin{property}\label{prop:gmr-tree}
Given a query $\mathcal{Q}$ and a node $N_i$, the signatures of $\mathcal{Q}$ and $N_i$ are $SIG_{\mathcal{Q}}$ and $SIG_i$ respectively. If $SIG_{\mathcal{Q}} = SIG_{\mathcal{Q}} \bigwedge SIG_i$, that means the query $\mathcal{Q}$ contains some same semantic conceptions as the objects in $N_i$, in other words, the query may be similar to some objects in $N_i$ on semantic level. Otherwise, $\mathcal{Q}$ may be dissimilar to the objects in the node.
\end{property}

\subsection{$k$NN Geo-Multimedia Cross-modal Search Algorithm}
Based on GMR-Tree and its property, we design an efficient spatial search algorithm to support $k$NN geo-multimedia cross-modal retrieval. The pseudo-code of $k$GMCMS algorithm is demonstrated in Algorithm~\ref{alg:kgmcms}. Algorithm~\ref{alg:nn} is the nearest neighbor search algorithm based on GMR-Tree, which is used in $k$GMCMS.

For Algorithm~\ref{alg:kgmcms}, in the first step, a priority queue $\mathcal{L}$ is initialized as a empty set and an integer $\alpha$ which is used for counting during the search. $\mathcal{R}$ is the set of results. First the algorithm puts the root node of GMR-Tree $\mathcal{G}$ into $\mathcal{L}$, and then calculates the signature for query $\mathcal{Q}$. In this calculation process, each element of semantic representation vector $\mathcal{Q}.M_T$ is reassigned by a hashing function $\mathcal{H}_{SIG}(.)$ which converts the element of $\mathcal{Q}.M_T$ into a hash code. After that, the search process is implemented by a \emph{While} loop. In this process, the nearest neighbor $o$ of query $\mathcal{Q}$ is found out and then the score of $o$ is calculated by score function $\mathcal{F}_{score}(\mathcal{Q},o)$ which is introduced in section~\ref{preliminary}. Here we set $\mu = 0.5$. That means the geographical distance proximity is same important as semantic correlation.

For Algorithm~\ref{alg:nn}, we initialize a variable $\mathcal{E}$ to store a node of GMR-Tree. $\mathcal{L}$ will be checked circularly if it is empty or not. If $\mathcal{L}$ is not empty, the algorithm gets a node stored in $\mathcal{L}$ by a $Dequeue(.)$ operation and put it into $\mathcal{E}$. If this node is a non-leaf node, and exist an object whose the $SIG$ matches the query, then measures the distance between $\mathcal{Q}$ and $MBR$ of $\mathcal{E}$. It will be put into $\mathcal{L}$ again. If $\mathcal{E}$ is a leaf node, all objects in it will be checked and put the object which matches the query in to $\mathcal{L}$.

\begin{algorithm}
\begin{algorithmic}[1]
\footnotesize
\caption{\bf $k$NN Geo-Multimedia Cross-Modal Search ($k$GMCMS)}
\label{alg:kgmcms}

\INPUT  A GMR-Tree $\mathcal{G}$, a query $\mathcal{Q}$.
\OUTPUT A results set $\mathcal{R}$.

\STATE Initializing: $\mathcal{R} \leftarrow \emptyset$;
\STATE Initializing: a priority queue $\mathcal{L} \leftarrow \emptyset$;
\STATE Initializing: an integer $\alpha \leftarrow 0$;

\STATE $\mathcal{L}.Enqueue(\mathcal{G}.Root,0)$
\FOR{each element $M_T^{(i)} \in \mathcal{Q}.M_T$}
    \STATE $M_T^{(i)} \leftarrow \mathcal{H}_{SIG}(M_T^{(i)})$;
\ENDFOR
\WHILE{$\alpha < \mathcal{Q}.k$}
    \STATE $PTR_o \leftarrow NearestNeighbor(\mathcal{Q}.\lambda,\mathcal{Q}.M_T,\mathcal{L})$
    \STATE $\mathfrak{o} \leftarrow LoadObject(PTR_o)$;
    \IF{$\mathcal{F}_{score}(\mathcal{Q},\mathfrak{o}) > \mathcal{F}_{score}(\mathcal{Q},\mathfrak{o}'),\forall \mathfrak{o}' \in \mathcal{O} \setminus \mathcal{R}$}
        \STATE $\mathcal{R} \leftarrow AddObject(\mathfrak{o})$;
        \STATE $\alpha \leftarrow \alpha+1$;
    \ENDIF
\ENDWHILE
\RETURN $\mathcal{R}$;
\end{algorithmic}
\end{algorithm}

\begin{algorithm}
\begin{algorithmic}[1]
\footnotesize
\caption{\bf NearestNeighbor($\mathcal{Q}.\lambda,\mathcal{Q}.M_T,\mathcal{L}$)}
\label{alg:nn}

\INPUT A query $\mathcal{Q}$, a list $\mathcal{L}$.
\OUTPUT A results set $\mathcal{R}$.

\STATE Initializing: a variable $\mathcal{E} \leftarrow \emptyset$;

\WHILE{$\mathcal{L}.IsNotEmpty()$}
    \STATE $\mathcal{E} \leftarrow \mathcal{L}.Dequeue()$;
    \IF{$\mathcal{E}$ is a non-leaf node}
        \FOR{each $<MBR,SIG,PTR_N>$ in $\mathcal{E}$}
            \IF{$SIG$ matches $\mathcal{Q}.M_T$}
                \STATE $\mathcal{L}.Enqueue(LoadNode(PTR_N),Dst(\mathcal{Q}.\lambda,MBR))$;
            \ENDIF
        \ENDFOR
    \ELSIF{$\mathcal{E}$ is a leaf node}
        \FOR{each $<MBR,SIG,PTR_o>$ in $\mathcal{E}$}
            \IF{$SIG$ matches $\mathcal{Q}.M_T$}
                \STATE $\mathcal{L}.Enqueue(LoadNode(PTR_o),Dst(\mathcal{Q}.\lambda,MBR))$;
            \ENDIF
        \ENDFOR
    \ELSE
        \RETURN $\mathcal{E}$;
    \ENDIF
\ENDWHILE

\end{algorithmic}
\end{algorithm}

\section{Experimental Evaluation}
\label{Experiment}

In this section, we conduct a comprehensive experiments on real and synthetic dataset to evaluate and efficiency of the proposed method in the paper, i.e., \textbf{DeCoSReS+GMR-Tree}.

\subsection{Experimental Settings}

\noindent\textbf{Workload.} A workload for $k$NN geo-multimedia cross-modal query experiment includes 100 input queries. The query locations are randomly selected from the locations of the underlying objects. By default, the number of final results $k=10$, and data number $N=80k$. We use response time and precision to evaluate the performance of the algorithms. The size of dataset is set to $40k$, $80k$, $120k$, $160k$ and $200k$. The number of results $k$ is set to 5, 10, 20, 50 and 100. Our experiments are run on a PC with Intel(R) CPU Xeon 2.60GHz and 16GB memory running Ubuntu 16.04 LTS Operation System. All algorithms in the experiments are implemented in Java.

\noindent\textbf{Dataset.} Our experiment aim to evaluate the performance of our solution on an real geo-multimedia dataset and synthetic dataset. The real dataset includes over one million geo-tagged images crawled from Flickr(http://www.flickr.com/), a popular Web site for users to share and embed personal photographs. we generate dataset SF by obtaining
the spatial locations from corresponding spatial datasets from Rtree-Portal (http://www.rtreeportal.org) and randomly geo-tagging these objects with images in ImageNet(http://image-net.org/index). ImageNet is a famous image database organized according to the WordNet hierarchy (currently only the nouns), in which each node of the hierarchy is depicted by hundreds and thousands of images. There are more than 100,000 synsets in WordNet, majority of them are nouns (80,000+). ImageNet provides on average 1000 images to illustrate each synset. Images of each concept are quality-controlled and human-annotated.

\noindent\textbf{Baseline.} To our best knowledge, we are the first to study the problem of $k$NN geo-multimedia cross-modal query. That means there are no existing approaches for this problem. We devise two baseline methods i.e., \textbf{DeCoSReS$+$R-Tree} and Semantic Matching~\cite{DBLP:conf/mm/RasiwasiaPCDLLV10}+R-Tree (\textbf{SM+R-Tree} for short) which is used SIFT+BoVW to extract the visual features. The geo-multimedia data such as geo-tagged images and geo-tagged texts are represented in the common semantic space.

\subsection{Experimental Results}

\subsubsection{Evaluation on Real Dataset}

\begin{figure*}
\newskip\subfigtoppskip \subfigtopskip = -0.1cm
\begin{minipage}[b]{0.99\linewidth}
\begin{center}
     \subfigure[{Different size of dataset}]{
     \includegraphics[width=0.48\linewidth]{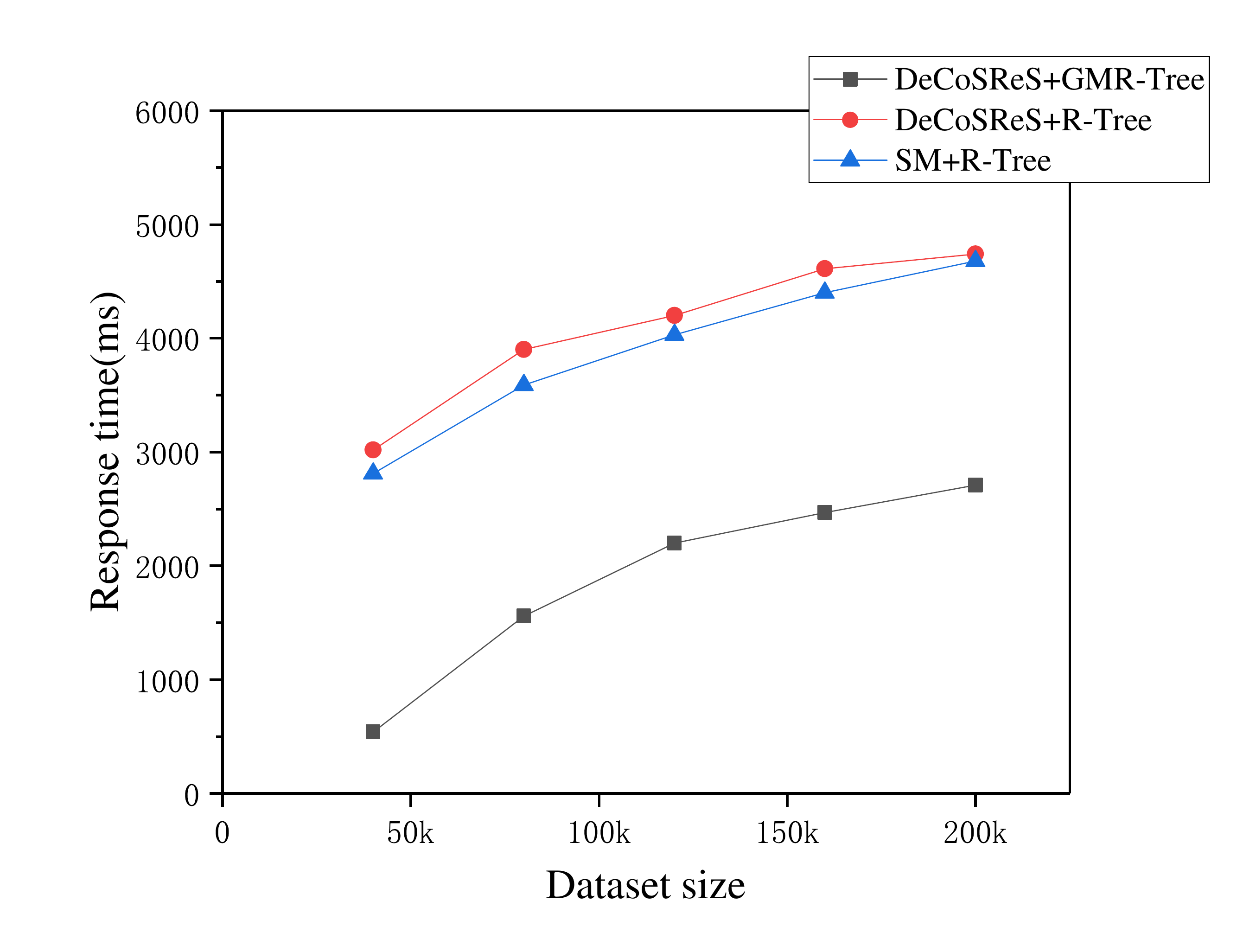}
     }
     \subfigure[{Different number of results}]{
     \includegraphics[width=0.48\linewidth]{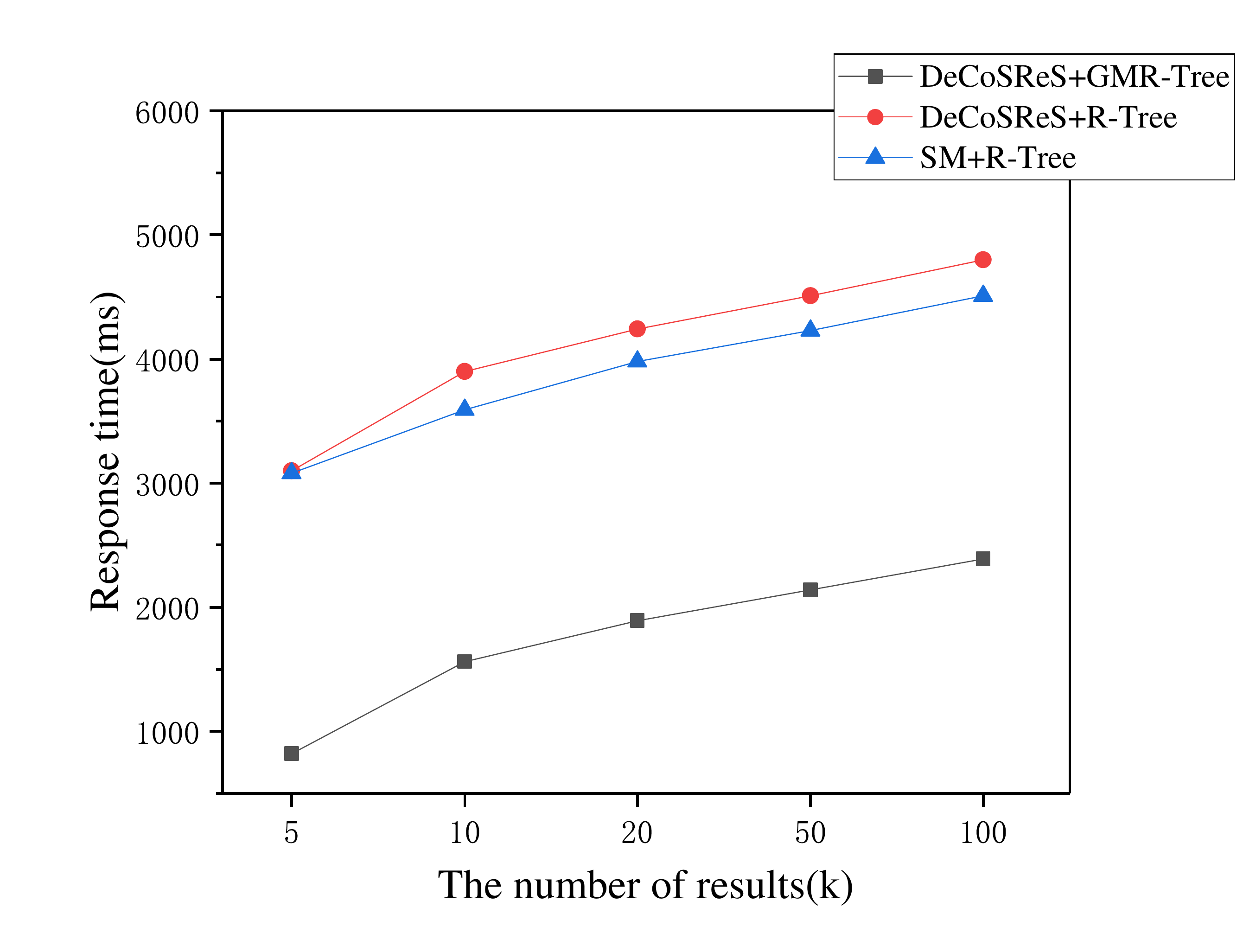}
     }
   \captionsetup{justification=centering}
       \vspace{-0.2cm}
\caption{Evaluation on real dataset}
\label{fig:fig6}
\end{center}
\end{minipage}
\label{fig:k}
\end{figure*}

\noindent\textbf{Evaluation on different size of dataset.}  We evaluate the performance of our approach, i.e., DeCoSReS+GMR-Tree and two baselines DeCoSReS$+$R-Tree and SM+R-Tree with the increment of dataset size. In Figure.~\ref{fig:fig6}(a) we can see that with the increasing of dataset size, the response time of all these methods increasing gradually. DeCosReS+GMR-Tree has the smallest response time due to the efficient indexing structure. It increase obviously and slow down when the dataset size is larger than $100k$. The performance of SM+R-Tree is a litter higher than DeCoSReS+R-Tree, and the latter is showing an rise trend of volatility between $50k$ and $200k$. And at last, the response time of these two baselines are nearly 5000ms.

\noindent\textbf{Evaluation on different number of results $k$.} We evaluate the performance of DeCoSReS+GMR-Tree, DeCoSReS$+$R-Tree and SM+R-Tree with the increasing of number of results $k$. In this evaluation, we increase $k$ from 5 to 100 in this experiment. Figure.~\ref{fig:fig6}(b) demonstrates that the response time of out method is going up with the rising of $k$. When $k = 5$, this response time is smaller than 1000ms, and it increase step by step in the interval of $[10,100]$. By contrast, the response time of DeCoSReS$+$R-Tree and SM+R-Tree are much higher than DeCoSReS+GMR-Tree. Likewise, they climb gradually with the rising of $k$. Similar to the situation shown in Figure.~\ref{fig:fig6}(a), the performance of the two baselines are similar.

\subsubsection{Evaluation on Synthetic Dataset}

\begin{figure*}
\newskip\subfigtoppskip \subfigtopskip = -0.1cm
\begin{minipage}[b]{0.99\linewidth}
\begin{center}
     \subfigure[{Different size of dataset}]{
     \includegraphics[width=0.48\linewidth]{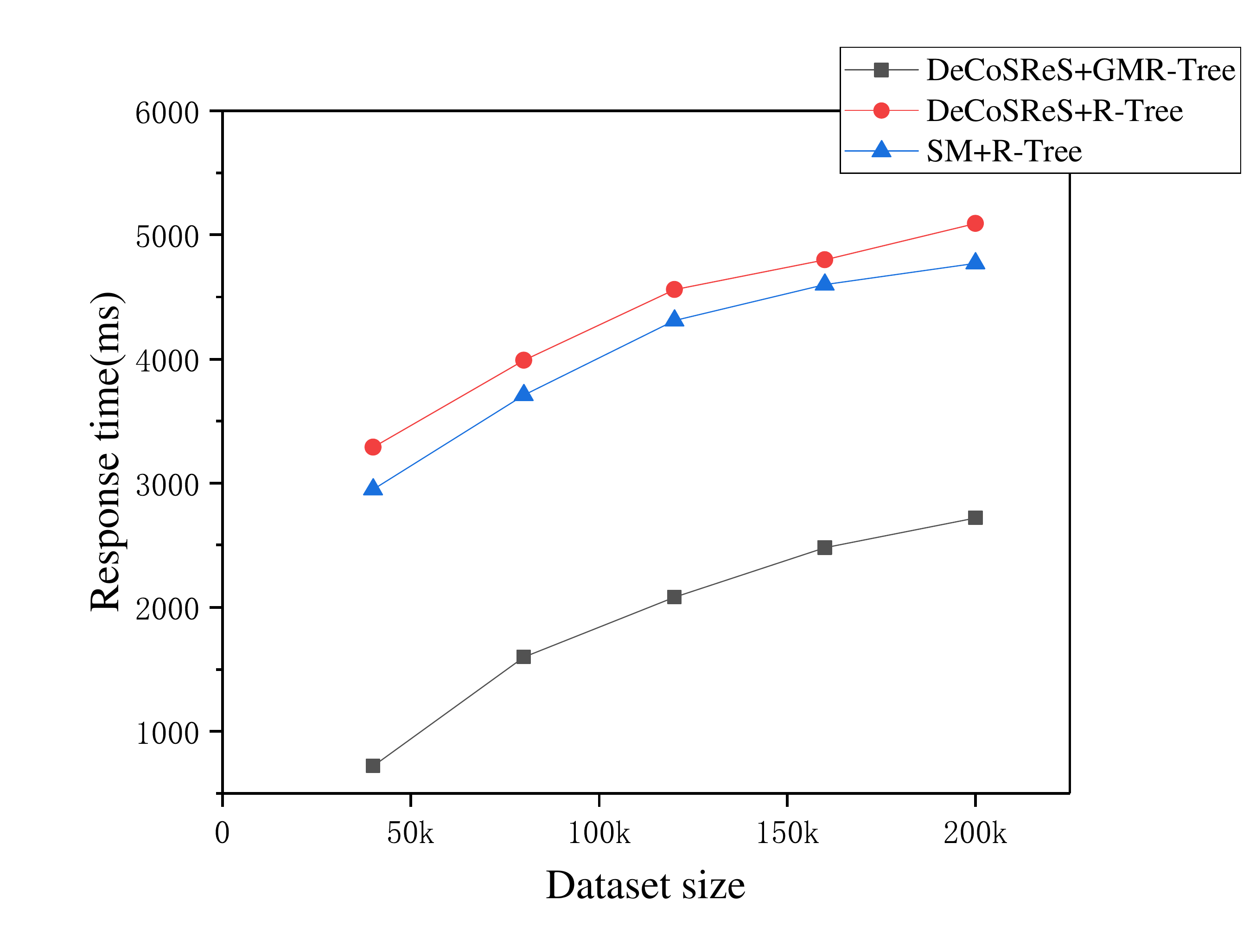}
     }
     \subfigure[{Different number of results}]{
     \includegraphics[width=0.48\linewidth]{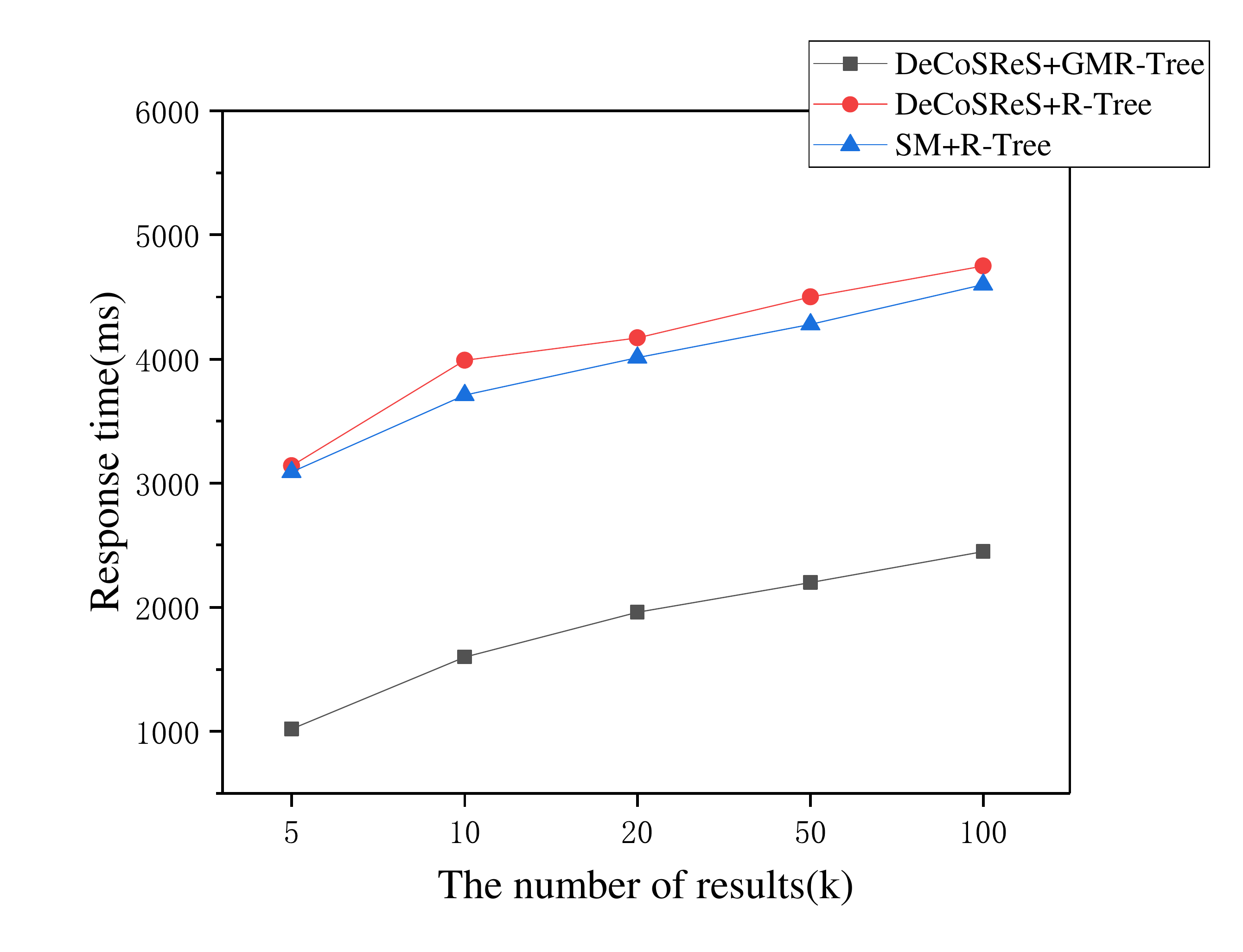}
     }
   \captionsetup{justification=centering}
       \vspace{-0.2cm}
\caption{Evaluation on synthetic dataset}
\label{fig:fig7}
\end{center}
\end{minipage}
\label{fig:k}
\end{figure*}

\noindent\textbf{Evaluation on different size of dataset.} On synthetic dataset, we evaluate the performance of these three methods on different size of dataset. Figure.~\ref{fig:fig7}(a) illustrates that the performance of DeCoSReS+GMR-Tree, DeCoSReS$+$R-Tree and SM+R-Tree decrease the step by step with the increasing of dataset size. By comparison, our method has obvious advantages in efficiency. When the dataset size is smaller than $100k$, the response time of it is less than 2000ms apparently. The time cost of DeCoSReS$+$R-Tree and SM+R-Tree is increasing faster in the interval of $[50k,100k]$ and after that, the growth of them are slow down.

\noindent\textbf{Evaluation on different number of results $k$.} Figure.~\ref{fig:fig7}(b) shows the evaluation of efficiency of our method and two baselines with the increment of number of results $k$. With $k$ increasing from 10 to 100, the efficiency of our method slows down litter by litter, and it is the most efficient approach in these three. The performance of DeCoSReS$+$R-Tree and SM+R-Tree are still similar, at $k=5$ they are nearly 3000ms and when $k=100$, they increase to 4500ms around. 

\subsubsection{Evaluation on retrieval precision on real dataset}

\begin{figure*}
\newskip\subfigtoppskip \subfigtopskip = -0.1cm
\begin{minipage}[b]{0.99\linewidth}
\begin{center}
     \subfigure[{The precision of DeCoSReS+GMR-Tree}]{
     \includegraphics[width=0.48\linewidth]{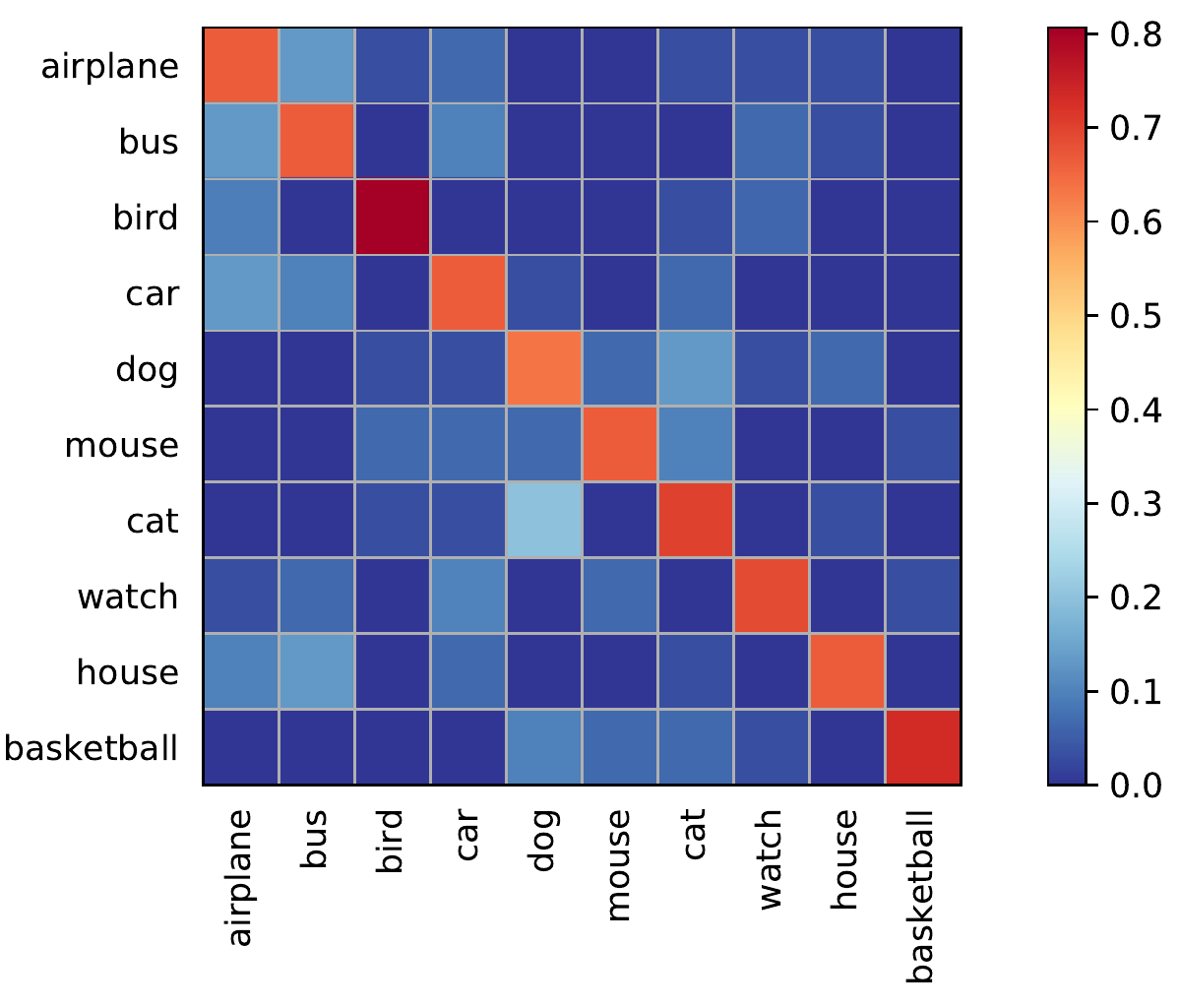}
     }
     \subfigure[{The precision of SM+R-Tree}]{
     \includegraphics[width=0.48\linewidth]{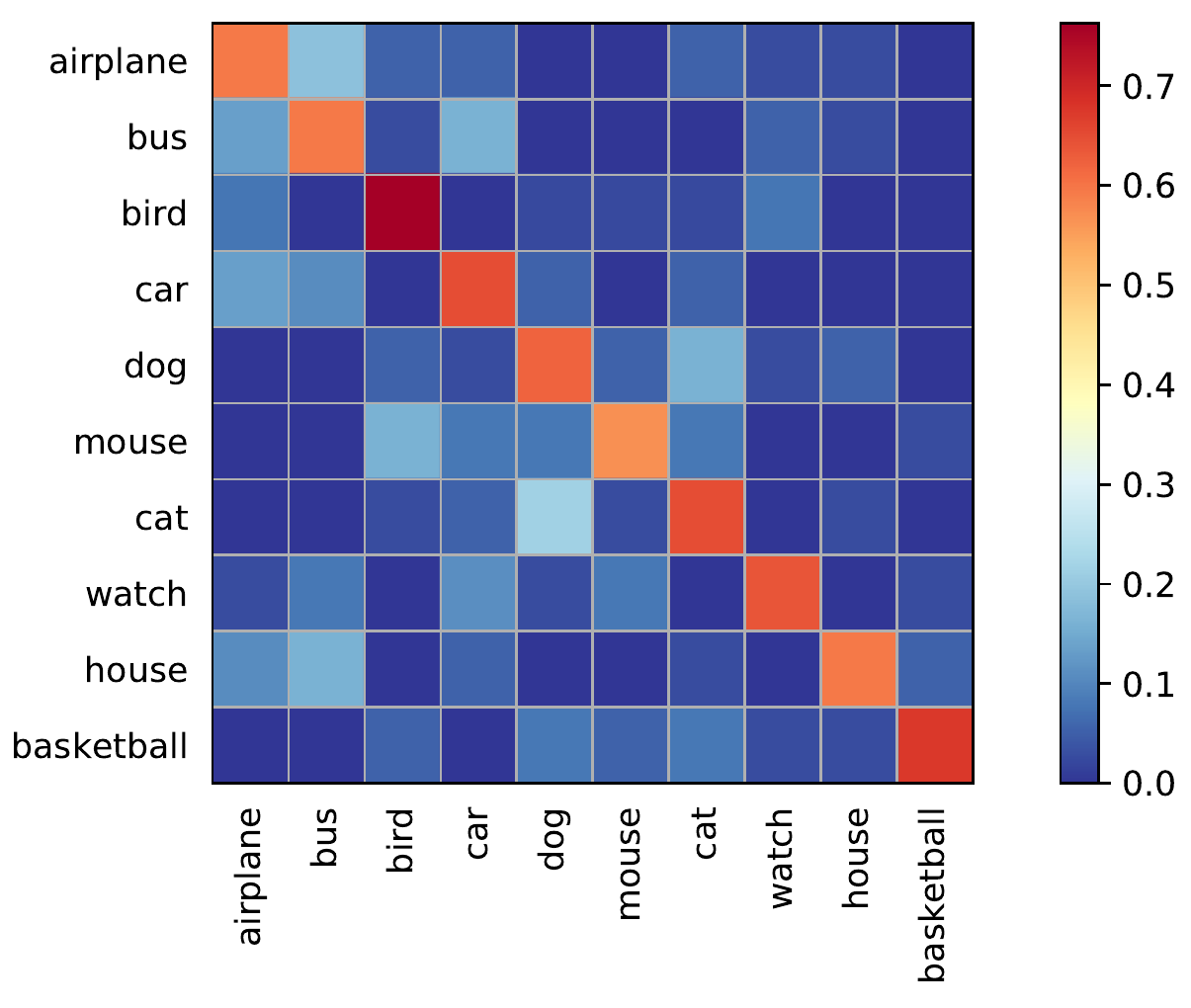}
     }
   \captionsetup{justification=centering}
       \vspace{-0.2cm}
\caption{Evaluation on real dataset}
\label{fig:fig8}
\end{center}
\end{minipage}
\label{fig:k}
\end{figure*}

Figure.~\ref{fig:fig8} demonstrates that the confusion matrices of retrieval on real dataset by methods DeCoSReS+GMR-Tree and SM+R-Tree. The techniques of semantic representation space construction is different, which is the main factor affecting the retrieval precision. It is easily to find that the precision of our method is a litter higher than SM+R-Tree, which is nearly 0.8. That means the visual features extracted by CNN are more discriminative than the features extracted by SITF+BoVW. 
\section{Conclusion}
\label{Conclusion}

In this paper, we propose a novel problem named $k$NN geo-multimedia cross-modal retrieval which aims to return $k$ nearest geo-multimedia objects which are highly similar to the query in the aspect of semantics. To solve this problem, we at first time propose the definition of geo-multimedia object and $k$NN geo-multimedia cross-modal query, as well as cross-modal semantic representation space. To address the ticklish problem of semantic gap between different modalities, we present an approach called  cross-modal semantic matching and a framework via deep learning techniques to construct a common semantic representation space for multi-modal data. To implement efficient spatial search, we propose a novel hybrid index structure name GMR-Tree which is a combination of geo-multimedia data and R-Tree. Based on it, we design a efficient $k$NN search algorithm named kGMCMS to support geo-multimedia cross-modal retireval. The experimental results illustrate that our solution outperform the-state-of-the-art methods.

\textbf{Acknowledgments:} This work was supported in part by the National Natural Science Foundation of China
(61702560, 61472450), the Key Research Program of Hunan Province(2016JC2018), project (2018JJ3691) of Science and Technology Plan of Hunan Province, and the Research and Innovation Project of Central South University Graduate Students(2018zzts177). 



\bibliographystyle{spmpsci}      

\bibliography{ref}

\end{document}